\documentclass[12pt]{article}

\pdfoutput=1

\usepackage{amssymb}
\usepackage{amsmath}
\usepackage{amssymb}
\usepackage{graphicx}
\usepackage{amsfonts}
\usepackage{hyperref}

\def\unit{\relax{\rm 1\kern-.26em I}}
\def\nada{\relax{\rm 0\kern-.30em l}}

\makeatletter
\renewcommand\section{\@startsection {section}{1}{\z@}%
                                 {-3.5ex \@plus -1ex \@minus -.2ex}
                                   {2.3ex \@plus.2ex}%
                                   {\normalfont\large\bfseries}}
\renewcommand\subsection{\@startsection{subsection}{2}{\z@}%
                                   {-3.25ex\@plus -1ex \@minus -.2ex}%
                                     {1.5ex \@plus .2ex}%
                                     {\normalfont\bfseries}}
\renewcommand\subsubsection{\@startsection{subsubsection}{3}{\z@}%
                                   {-3.25ex\@plus -1ex \@minus -.2ex}%
                                     {1.5ex \@plus .2ex}%
                                     {\normalfont\itshape}}
\makeatother

\newcommand{\be}{\begin{equation}}
\newcommand{\ee}{\end{equation}}
\newcommand{\bea}{\begin{eqnarray}}
\newcommand{\eea}{\end{eqnarray}}
\newcommand{\barr}{\begin{array}}
\newcommand{\earr}{\end{array}}
\newcommand {\lessim} {\ {\raise-.5ex\hbox{$\buildrel<\over\sim$}}\ } 

\def\beq{\begin{equation}}
\def\eeq{\end{equation}}
\def\be{\begin{equation}}
\def\ee{\end{equation}}
\def\bea{\begin{eqnarray}}
\def\eea{\end{eqnarray}}

\DeclareRobustCommand{\SkipTocEntry}[4]{}

\begin{document}

\begin{titlepage}

\setcounter{page}{1} \baselineskip=19.5pt \thispagestyle{empty}

\begin{flushright}
\end{flushright}
\vfil

\begin{center}
{\LARGE Resonant non-Gaussianity with equilateral properties}

\end{center}
\bigskip\

\begin{center}
{\large Rhiannon Gwyn$^\#$, Markus Rummel$^{\ast}$ and Alexander Westphal$^\dag$}
\end{center}

\begin{center}
\textit{$^\#$AEI Max-Planck-Institut f\"ur Gravitationsphysik, D-14476 Potsdam, Germany}\\
\textit{$^\ast$II. Institut f\"ur Theoretische Physik der Universit\"at Hamburg, D-22761 Hamburg, Germany}\\
\textit{$^\dag$Deutsches Elektronen-Synchrotron DESY, Theory Group, D-22603 Hamburg, Germany}\\

\end{center} \vfil

\noindent We discuss the effect of superimposing multiple sources of resonant non-Gaussianity, which arise for instance in models of axion inflation. The resulting sum of oscillating shape contributions can be used to ``Fourier synthesize" different non-oscillating shapes in the bispectrum. As an example we reproduce an approximately equilateral shape from the superposition of ${\cal O}(10)$ oscillatory contributions with resonant shape.  This implies a possible degeneracy between the equilateral-type non-Gaussianity typical of models with non-canonical kinetic terms, such as DBI inflation, and an equilateral-type shape arising from a superposition of resonant-type contributions in theories with canonical kinetic terms. The absence of oscillations in the 2-point function together with the structure of resonant N-point functions give a constraint of $f_{NL} \lessim \mathcal{O}(5)$ for equilateral non-Gaussianity with resonant origin, but this constraint can be avoided when additional
U(1)s are involved in the breaking of the shift symmetry. We comment on the questions arising from possible embeddings of this idea in a string theory setting.

\vfil
\begin{flushleft}
\today
\end{flushleft}

\end{titlepage}

\newpage
\tableofcontents
\newpage

\section{Introduction}\label{intro}

Recent years have seen the advent of precise observations of the cosmic microwave background radiation (CMB)~\cite{Komatsu:2010fb,Dunkley:2010ge,Keisler:2011aw}, and the redshift-distance relation for large samples of distant type IA supernovae~\cite{Riess:1998cb,Perlmutter:1998np}, as well as a host of additional increasingly precise measurements such as  baryon acoustic oscillations~\cite{Percival:2009xn} or the determination of the Hubble parameter $H_0$ by the Hubble Space Telescope key project~\cite{Riess:2011yx}. These results are so far in concordance with a Universe very close to being spatially flat, as well as with a pattern of coherent acoustic oscillations in the early dense plasma which was seeded by an almost scale-invariant spectrum of super-horizon size curvature perturbations with Gaussian distribution. This structure is a direct consequence of a wide class of models of cosmological inflation driven by the potential energy of a single scalar field with canonically normalized kinetic 
energy 
in (strict) slow-roll. In these models of inflation the spectrum of scalar curvature perturbations generated during the quasi-de-Sitter phase is necessarily Gaussian, with 3-point interactions generating very small non-Gaussian deviations $f_{NL}={\cal O}(\epsilon,\eta)$ of order of the slow-roll parameters during inflation~\cite{Maldacena:2002vr}. Consequently there are large classes of inflationary models which generate sizable levels of non-Gaussianity and are characterized by departures from one or more of the defining properties of canonically normalized single-field slow-roll inflation. 

Non-Gaussianity can arise from a non-canonical kinetic term for a single scalar field characterized by the presence of higher-derivative terms, from controlled non-permanent violations of slow-roll  such that the overall inflationary behaviour is preserved, or from the presence of several light scalar fields during inflation. A series of results have established that these different mechanisms of generating a large 3-point function and its associated non-Gaussianity lead to different `shapes', the distributions of the amount of non-Gaussianity over the momentum-conservation dictated triangular domain of the three fluctuation momenta in the 3-point function. These shape functions allow for discrimination among many different classes of inflationary models, provided one has sufficient measurement accuracy for the shape, and one has shown the existence of a complete orthonormal system of shapes of non-Gaussianity in the space of inflation models. For two recent reviews see e.g.~\cite{Chen:2010xka,Koyama:2010xj}.

In this article, we explore the degeneracy, at the level of non-Gaussianity, of different classes of inflationary models. Specifically we study degeneracy between models of inflation characterized by so-called non-canonical kinetic terms  such as DBI inflation \cite{Silverstein:2003hf, Alishahiha:2004eh} and theories which are purely canonical. By non-canonical kinetic terms we mean terms of the form $X^n$, where $X = \frac12\partial^\mu\phi\partial_\mu\phi$, so that a non-canonical Lagrangian is given by some general $P(X, \phi)$ \cite{Franche:2009gk}. These are thus a subset of the theories described by the Horndeski action, which is the most general action for a single field with at most two time derivatives acting on it at the level of the equation of motion, i.e. the most general action for a single scalar field without ghosts or Ostrogradski instabilities (See e.g.~\cite{RenauxPetel:2011sb} and \cite{Ribeiro:2012ar}). 

Theories in this class give rise to an equilateral-type non-Gaussianity inversely proportional to the reduced speed of sound, $c_s$ (see e.g. \cite{Chen:2006nt}):
\begin{eqnarray}
f_{NL}^{equil} & \sim & \frac{1}{c_s^2},
\end{eqnarray}
where
\begin{eqnarray}
c_s^2 & = & \left (1 + 2X \frac{P_{XX}}{P_X} \right)^{-1}.
\end{eqnarray}

By contrast, canonical slow-roll single scalar field models give rise to Gaussian spectra (with corrections of order the slow-roll parameters) for which all odd-point correlation functions vanish and all even-point correlations are given in terms of the two-point functions. How then could there be any observational degeneracy between a model of non-canonical inflation and a canonical single scalar field model of inflation? In fact, single scalar field models with canonical kinetic term give rise to (approximately) Gaussian spectra only as long as the slow-roll conditions are always satisfied.   As argued in \cite{Chen:2008wn}, these theories can generate observably large non-Gaussianities when there are features in the potential. The two classes of such models considered in \cite{Chen:2008wn} are: 1) potentials with steps resulting in a localized violation of slow-roll, and 2) potentials with a small modulation giving rise to oscillating slow-roll parameters which can induce resonance in the three-point 
function of the quantum fluctuations. This gives an enhanced non-Gaussianity with a specific shape, termed resonant non-Gaussianity \cite{Chen:2008wn} which is orthogonal to the local and equilateral shapes \cite{Flauger:2010ja}. Note that in this case the three-point function can be very large while the two-point function remains small. Resonant non-Gaussianity comes with the additional property of allowing for very efficient calculability of the $N$-point functions up to $N=10\ldots 20$~\cite{Leblond:2010yq}.

However, as we shall show, a closely related model, in which a sum of such modulations in the potential is present, leads to a very different non-Gaussianity. Its shape is of the equilateral type in the sense that its momentum average has a large overlap with the standard equilateral one, despite the lack of non-canonical kinetic terms. Yet, periodic scale dependent features remain. Thus observation of equilateral non-Gaussianity might not necessarily imply that non-canonical kinetic terms were present or relevant in the underlying model of inflation.  This degeneracy at the level of the three-point function leads us to conclude that it may not be as easy as previously hoped to distinguish amongst single-field models of inflation.\footnote{Note that this degeneracy can also arise via a different mechanism, as  in \cite{Barnaby:2010vf, Barnaby:2011qe}. Here, equilateral non-Gaussianity can arise from inflaton fluctuations sourced by gauge quanta via the pseudoscalar coupling $\phi F \tilde F$ to some gauge 
field. However, the non-Gaussianity we consider has a different origin, coming only from the potential for $\phi$.We thank Neil Barnaby for bringing this work to our attention.}

The non-observation of oscillating contributions to the 2-point function, and the structure of the resonant $N$-point functions can place tight constraints on the maximum amount of resonantly generated non-Gaussianity with equilateral characteristics. Given such a constraint from the power spectrum, equilateral non-Gaussianity with a resonant origin would be restricted to  $f_{NL}\sim {\cal O}(5)$ or lower~\cite{Behbahani:2011it}. Then an observation of $f_{NL}^{equil}$ greater than this would rule out a resonant origin. In the case of non-observation of non-Gaussianity by PLANCK, one would have to await future 21cm observations, which may achieve a detection capability of $f_{NL}\gtrsim {\cal O}(0.01)$ down to the slow-roll level itself due to the extremely large number of modes (${\cal O}(10^{16})$) available~\cite{Loeb:2003ya,Cooray:2006km}. For equilateral-type non-Gaussianities of this magnitude there would then be a degeneracy between a possible non-canonical and a resonant origin. However, the power 
spectrum constraint can be avoided in resonant models in which symmetry under time translation is collectively broken.\footnote{We thank Daniel Green and Siavosh R. Behbahani for pointing this out to us.} In this case, it is possible to have large resonant non-Gaussianity without a large signal in the power spectrum \cite{Behbahani:2012be}.

\section{A sum of resonant bispectra}

\subsection{Potential, solution and slow-roll parameters}
\begin{figure}[htp]
\centerline{\includegraphics[scale=.8]{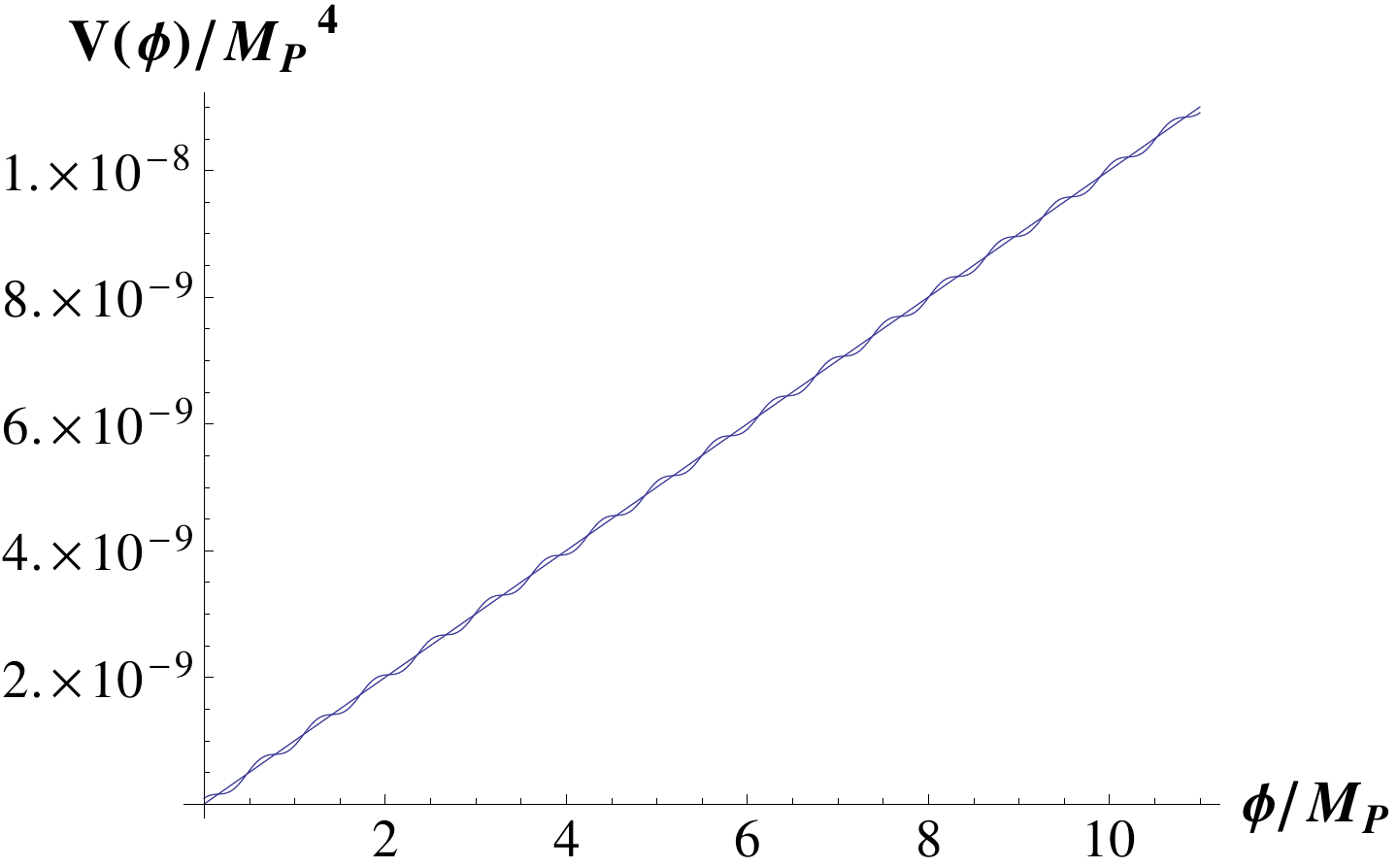}}
\caption{Modulated linear potential, for a convenient choice of  ${c_i, f_i}$  values.}
\label{ResPot}
\end{figure}

From an effective field theory point of view, we can begin with a model of a single scalar field with canonical kinetic term and modulated potential 
\be
\label{sum_pot}
V(\phi) = V_0 (\phi) + \sum_i A_i  \cos \left (\frac{\phi + c_i}{f_i} \right),
\ee
where we have generalized the modulated potential in  \cite{Chen:2008wn, Flauger:2010ja}  to the case where the modulation is a series of terms with varying phases. For suitable coefficients and values of $f_i$, the sum remains a small perturbation on the potential $V_0(\phi)$ (see Figure~\ref{ResPot}). We will show that slow-roll inflation for this theory is supported for sufficiently large $f_i$, while a large variation in the slow-roll parameters is possible. In Sections \ref{pwrspectrum} and \ref{thebispectrum} we give the power spectrum and bispectrum respectively, finding that the bispectrum is given by a series of the resonant bispectra found for a singly modulated potential in \cite{Chen:2008wn, Flauger:2010ja}. This bispectrum was first calculated in \cite{Chen:2008wn}; here we follow closely the notation and analysis in \cite{Flauger:2010ja}. The details of our calculations are given in Appendix~\ref{computation}.

The equation of motion for the inflaton $\phi$ with potential eq.~\eqref{sum_pot} is 

\be
\ddot \phi + 3 H \dot \phi + V_0'(\phi) =  \sum_i \frac{A_i}{f_i} \sin \left ( \frac{\phi + c_i}{f_i}\right ).
\ee

As in \cite{Flauger:2010ja} we treat the oscillations as a perturbation and write $\phi = \phi_0 + \phi_1 + ...$ where $\phi_0$ is the solution in the absence of modulations, given by 
\be
\dot \phi_0 = - \frac{V_0'(\phi_0)}{\sqrt{3 V_0(\phi_0)}}.
\ee

The equation of motion for $\phi_1$, written in terms of derivatives with respect to $\phi_0$ (denoted by a prime) is 
\be
\phi_1'' - \frac{3}{\sqrt{2 \epsilon_{V_0}}} \phi_1' - \left (\frac{3}{2} - \frac{3 \eta_{V_0}}{2 \epsilon_{V_0}}  \right )\phi_1 = \frac{3}{2 \epsilon_{V_0}V_0}  \sum_i \frac{A_i}{f_i} \sin \left ( \frac{\phi_0 + c_i}{f_i}\right ) ,
\ee
where $V_0 = V_0(\phi_0)$, and $\epsilon_{V_0}$ and $\eta_{V_0}$ are the canonical slow-roll parameters evaluated for the potential $V_0$. Note that $\epsilon'_{V_0} = \sqrt{2 \epsilon_{V_0}} (\eta_{V_0} - 2 \epsilon_{V_0})$. The solution is a generalization of that found in \cite{Flauger:2010ja}; to first order in slow-roll parameters and assuming $f_i \ll \sqrt{2 \epsilon_\star} \, \forall \, i$, it is given by
\be
\phi_1(t)  =  - \frac{3}{2 \epsilon_\star V_0(\phi_\star)} \sum_i A_i f_i \sin \left ( \frac{\phi_0 + c_i}{f_i}\right ) ,
\ee
where the subscript $\star$ indicates that $\epsilon = \epsilon_{V_0}(\phi_\star)$ is evaluated at horizon exit. This solution works because $\phi_1 = \sum \phi_i$, where each $\phi_i$ satisfies the EOM for a specific $f_i$.  $\phi_0(t)$ is time dependent but, for the period of interest, in which modes observed today exit the horizon, we can approximate it as being given by the value at horizon crossing, i.e. $\phi_\star$, except in the oscillating term. In other words we have assumed that the dominant variation in $\phi_1$ is due to the modulating terms. Defining 
\be
\label{bidef}
b_i ^\star \equiv  \frac{A_i}{V_0'(\phi_\star) f_i},
\ee
we have that the full solution is well approximated by
\be
\label{phitsoln}
\phi(t)  = \phi_0(t) - \sum_i \frac{3 b_i^\star f_i^2}{\sqrt{2 \epsilon_\star}} \sin  \left ( \frac{\phi_0 + c_i}{f_i}\right ) .
\ee

The slow-roll parameters $\epsilon = - \frac{\dot H}{H^2}$ and $\delta = \frac{\ddot H}{2 H \dot H}$ can be calculated for the potential eq.~\eqref{sum_pot} and are found to be:
\begin{align}
\begin{aligned}
\epsilon & = \epsilon_\star + \left ( \frac{V_0'}{V_0}\right ) ^2 \phi_1',\\
\epsilon & = \epsilon_\star - 3 \sqrt{2 \epsilon_\star} \sum_i b^\star_i f_i \cos \left ( \frac{\phi_0 + c_i}{f_i}\right ) + \mathcal O((b^\star_i)^2), \\
\epsilon & = \epsilon_0 + \epsilon_1 +  \mathcal O((b^\star_i)^2;\\
\delta & = \epsilon_\star - \eta_\star - \sqrt{2 \epsilon_\star} \phi_1'' + \epsilon_\star ( \eta_\star - \epsilon_\star) \phi_1,\\
\label{delta1} \delta & = \epsilon_\star - \eta_\star - \sum_i 3 b^\star_i \sin   \left ( \frac{\phi_0 + c_i}{f_i}\right ) + \mathcal O((b\star^i)^2,\\
\delta &= \delta_0 + \delta_1 + \mathcal O((b^\star_i)^2.
\end{aligned}
\end{align}

The SR parameter of interest for observation is $\frac{\dot \delta_1}{H}$ where $\delta_0 = \epsilon_\star - \eta_\star$ is the value of $\delta$ coming from the unmodulated potential $V_0$. Resonance with the oscillating $\frac{\dot \delta}{H}$ is what is responsible for a large resonant non-Gaussianity in \cite{Flauger:2010ja}. Here, we find
\bea
\label{deltadot}
\frac{\dot \delta _1}{H} & = & \sum_i \frac{\sqrt{2 \epsilon_\star}}{f_i} 3 b^\star_i \cos  \left ( \frac{\phi_0 + c_i}{f_i}\right ). 
\eea
Note that $\frac{\sqrt{2 \epsilon_\star}}{f_i}$ is large for all $i$ for the solution eq.~\eqref{phitsoln}. 

\subsection{The power spectrum}
\label{pwrspectrum}
As in \cite{Flauger:2010ja}, the Mukhanov-Sasaki equation for the mode function ${\mathcal R_k}$ of the curvature perturbation in our case is given by (see the Appendix for details)
\be
\frac{d^2 \mathcal R_k} {d x^2} - \frac{2(1 +\delta_1(x))}{x} \frac{d \mathcal R_k }{dx} + \mathcal R_k = 0,
\ee
where $x = - k \tau$ for $\tau$ the conformal time. We have neglected $\epsilon_0$ and $\delta_0$ and are working only to leading order in $\epsilon_\star$. The form of this equation is the same as in \cite{Flauger:2010ja}, but $\delta_1$ is now defined in eq.~\eqref{delta1} and given by
\begin{equation}
 \delta_1 = - 3 \sum_i b_i^\star \sin \left (\frac{\phi_0 + c_i}{f_i} \right ).
\end{equation}

We will find a sum of integrals each encountering resonance at a value $x_{{\rm res}, i}$. As in the case of a single modulation, the effect of $\delta_1$ is negligible both at early times $x \gg x_{{\rm res},i}$ and at late times $x \ll x_{{\rm res}, i}$. We therefore look for a solution of the form 

\be
\mathcal R_k (x) = \mathcal R_{k,0}^{(o)} \left [i \sqrt{\frac{\pi}{2}} x^{3/2} H_{3/2}^{(1)} (x) - c_k^{(-)}(x) i \sqrt{\frac{\pi}{2}} x^{3/2} H_{3/2}^{(2)} (x)  \right ] ,
\ee
where $\mathcal R_{k,0}^{(o)}$ is the value of $\mathcal R_k (x)$ outside the horizon in the absence of modulations,  $H_{3/2}^{(1,2)} (x)$ are Hankel functions and $c_k^{(-)}(x)$ vanishes at early times. Then the Mukhanov-Sasaki equation gives an equation governing the time evolution of $c_k^{(-)}(x)$ which is given by (for $f_i \ll \sqrt{2 \epsilon_\star}$ for all $i$)
\be
\label{eqnforcminus2}
\frac{d}{dx} \left [e^{-2 i x} \frac{d}{dx}  c_k^{(-)}(x) \right ]  =  - 2 i \frac{\delta_1(x)}{x}.
\ee

To solve this we need $\delta_1$ as a function of the conformal time $\tau$. In terms of $\tau$ the background equation of motion is given by 
\be
\frac{d \phi_0}{d \ln( - \tau)} = \sqrt{2 \epsilon_\star},
\ee
with solution $\phi_0(\tau) = \phi_\star + \sqrt{2 \epsilon_\star} \ln \frac{\tau}{\tau_\star}$. This can be written as $\phi_0(x) = \phi_k + \sqrt{2 \epsilon_\star} \ln x$, where $\phi_k = \phi_\star - \sqrt{2 \epsilon_\star} \ln \frac{k}{k_\star}$ is the value of the field when the mode with comoving momentum $k$ exits the horizon.  Then, writing $c_k^{(-)}(x) = \sum_i c_{k,i}^{(-)}(x)$, we need to solve 
\be
\label{ckminusderiv}\frac{d}{dx} c_{k,i}^{(-)} (x) = - \frac{6 i b_i^\star f_i}{\sqrt{2 \epsilon_\star}} e^{2 i x} \cos  (\frac{\phi_k}{f_i} + \frac{c_i}{f_i} + \frac{\sqrt{2 \epsilon_\star}}{f_i} \ln x ). 
\ee
Thus each $c_{k,i}^{(-)}(x)$ behaves like $c_k^{(-)}(x)$ in \cite{Flauger:2010ja} but with $b^\star = b_\star^i, f = f_i$ and resonance at different $x$ values $x_{{\rm res},i} = (2 f_i \phi_\star)^{-1}$. 

 For each term we can use the stationary phase approximation to find 
\bea
\label{ckminus}c_{k,i}^{(-)} &= &3 b_i^\star \sqrt{\frac{\pi}{2}} \left ( \frac{f_i}{\sqrt{2 \epsilon_\star}}\right ) ^{1/2} e^{- i(\frac{\phi_k + c_i}{f_i})} ,\\
\Rightarrow \mathcal R_k (x) & = & \mathcal R_{k,0}^{(o)} \big [ i \sqrt{\frac{\pi}{2}} x^{3/2} H_{3/2}^{(1)} (x) - \sum_i \frac{3i  b_i^\star \pi}{2}\sqrt{ \frac{f_i}{\sqrt{2 \epsilon_\star}}} e^{- i(\frac{\phi_k + c_i}{f_i})} x^{3/2} H_{3/2}^{(2)} (x)\big ] .
\eea

Outside the horizon, $k/aH \ll 1$ i.e. $ x = - k \tau = k/aH \ll 1$ (using the definition of conformal time and the fact that during inflation $H$ is constant). Then we find
\begin{align}
\begin{aligned}
\mathcal R_k (x) = R_{k,0}^{(0)} &\left [ 1 + \sum_i 3 b_i^\star \sqrt{\frac{f_i \pi}{2\sqrt{2 \epsilon_\star}}}  \cos \left (\frac{\phi_k + c_i}{f_i} \right ) \right. \\ &- \left. i \sum_i 3 b_i^\star \sqrt{\frac{f_i \pi}{2\sqrt{2 \epsilon_\star}}}  \sin \left (\frac{\phi_k + c_i}{f_i} \right )  + \mathcal O({x^3})\right ]\,,
\label{mode_function_late}
\end{aligned}
\end{align}
and
\begin{equation}
|\mathcal R_k^{(0)} |^2 = |\mathcal R_{k,0}^{(0)}|^2 \left [1 + \delta n_s \right ]\,, 
\end{equation}
with
\begin{equation}
\label{deltans}
 \delta n_s = \sum_i 3 b_i^\star \left (\frac{2 \pi f_i}{\sqrt{2 \epsilon_\star}} \right )^{1/2} \cos \left (\frac{\phi_k + c_i}{f_i}\right)\,,
\end{equation}
which is just the generalization of (2.30) in \cite{Flauger:2010ja} one might have expected.

\subsection{The bispectrum}
\label{thebispectrum}
Next we ask if this generalization carries through to the bispectrum as calculated in \cite{Chen:2008wn, Flauger:2010ja}. As in the case of a single modulation, the leading contribution comes from 
\be
\label{bispectrum_int}
H_I(t) \supset - \int d^3 x a^3 (t) \epsilon(t) \dot \delta (t) \mathcal R^2 ({\bf x}, t)\dot {\mathcal R} ({\bf x}, t), 
\ee
where to linear order in $ b_i^\star$ we can use the approximations $\epsilon \approx \epsilon_\star, \delta \approx \delta_1$ and use the unperturbed mode functions\footnote{To see why we can use the unperturbed mode functions, see Appendix~\ref{unperturbed}.}
\be
\mathcal R_k (x) = \mathcal R_{k,0}^{(o)} i \sqrt{\frac{\pi}{2}} x^{3/2} H_{3/2}^{(1)} (x).
\ee


 The analysis of \cite{Flauger:2010ja} therefore carries over to our case, with the three-point function given by 
\be
<\mathcal R ({\bf k_1}, t) \mathcal R ({\bf k_2}, t) \mathcal R({\bf k_3}, t)> = (2 \pi)^7 \Delta_{\mathcal R}^4 \frac{1}{k_1^2 k_2^2 k_3^2}\delta^{3} ({\bf k_1} + {\bf k_2} + {\bf k_3}) 
\frac{\mathcal G (k_1, k_2, k_3)}{k_1 k_2 k_3}\,. \label{threepointGgen}
\ee
for
\be
\frac{\mathcal G (k_1, k_2, k_3)}{k_1 k_2 k_3} =  \frac{1}{8} \int_0^\infty dX \frac{\dot \delta_1}{H} e^{- i X} \left [ - i - \frac{1}{X} \sum_{i \neq j} \frac{k_i}{ k_j} + \frac{i}{X^2} \frac{K (k_1^2 + k_2^2 + k_3^2)}{k_1 k_2k_3} \right ] + c.c,
\ee
where $K \equiv k_1 + k_2 + k_3$, $X \equiv - K \tau$, and $\dot \delta_1$ is given by eq.~\eqref{deltadot}.  Equivalently this can be written
\be
\label{Gsum}
\frac{\mathcal G (k_1, k_2, k_3)}{k_1 k_2 k_3} =  \frac{1}{8} \sum_i \int_0^\infty dX \frac{\dot \delta_{1,i}}{H} e^{- i X} \left [ - i - \frac{1}{X} \sum_{i \neq j} \frac{k_i}{ k_j} + \frac{i}{X^2} \frac{K (k_1^2 + k_2^2 + k_3^2)}{k_1 k_2k_3} \right ] + c.c.
\ee
where 
\be
\frac{\dot \delta _{1,i}}{H}  \,\, = \,\, \frac{\sqrt{2 \epsilon_\star}}{f_i} 3 b^\star_i \cos  \left ( \frac{\phi_0 + c_i}{f_i}\right ) \,\, = \, \,   \frac{\sqrt{2 \epsilon_\star}}{f_i} 3 b^\star_i \cos  \left ( \frac{\phi_K + \sqrt{2 \epsilon_\star} \ln X + c_i}{f_i}\right ).
\ee
Now, for each integral in the sum the biggest contribution comes from $X_{{\rm res}, i} = \frac{\sqrt{2 \epsilon_\star}}{f_i}$. This can be seen by defining
\begin{align}
\begin{aligned}
\mathcal I_{K,i} &= \frac{3 b_i^\star \sqrt{ \epsilon_\star} }{ f_i}\int_0 ^\infty d X e^{- i X} \cos  \left ( \frac{\phi_K + \sqrt{2 \epsilon_\star} \ln X + c_i}{f_i}\right )\\
&= \frac{3 b_i^\star \sqrt{2 \epsilon_\star} }{2 f_i} \int_0 ^\infty d X ( e^{i (\phi_K + c_i)/f_i}e^{- i( X -\frac{\sqrt{2 \epsilon_\star} \ln X}{f_i}) }  +e^{- i (\phi_K + c_i)/f_i} e^{ - i (X +\frac{\sqrt{2 \epsilon_\star} \ln X}{f_i})}).
\end{aligned}
\end{align}

The phase of the first term can be zero while the phase of the second cannot, which means that the first term will dominate, with its dominant contribution at $X_{{\rm res}, i} = \frac{\sqrt{2 \epsilon_\star}}{f_i}$. This confirms that the main contribution to the bispectrum arises when the modes are still deep in the horizon. In this regime, we can approximate $1/X$ and $1/X^2$ by $1/{X_{{\rm res}, i}}$ and $1/{X_{{\rm res}, i}^2}$ in eq.~\eqref{Gsum} and we have
\be
\frac{\mathcal G (k_1, k_2, k_3)}{k_1 k_2 k_3} = \frac{1}{4}\sum_i \left [{\rm Im} \mathcal I_{K,i} - \frac{1}{X_{{\rm res}, i}} \sum_{\ell \neq m} \frac{k_\ell}{ k_m} {\rm Re} \mathcal I_{K,i} - \frac{1}{X_{{\rm res}, i}^2}   \frac{K (k_1^2 + k_2^2 + k_3^2)}{k_1 k_2k_3} {\rm Im} \mathcal I_{K,i} \right ].
\ee
Further we can evaluate $\mathcal I_{K,i}$ in the stationary phase approximation to get
\be
\mathcal I_{K,i} = \frac{3 b_i^\star \sqrt{ 2 \pi } }{2}\left (  \frac{\sqrt{2 \epsilon_\star}}{f_i} \right ) ^{3/2}e^{i (\phi_K+c_i)/f_i}e^{ - \frac{i \pi}{4}} e^{- i (\frac{\sqrt{2 \epsilon_\star}}{f_i} - \frac{\sqrt{2 \epsilon_\star}}{f_i} \ln \frac{\sqrt{2 \epsilon_\star}}{f_i})}. \label{Ikifinal}
\ee

Using $ \phi_K = \phi_\star - \sqrt{2 \epsilon_\star} \ln \frac{K}{k_\star}$, this gives us the shape of the non-Gaussianity as a sum of the resonant non-Gaussianity shapes in \cite{Flauger:2010ja} (up to an overall phase):\footnote{Note that each term in the sum over $i$ in eq.~\eqref{sumbispectrum} actually has a different phase contribution which depends on the $f_i$, see eq.~\eqref{Ikifinal}. However, we can absorb these terms into the $c_i$ which we are free to choose, such that eq.~\eqref{sumbispectrum} is correct.}

\begin{align}
\begin{aligned}
 \label{sumbispectrum}
\frac{\mathcal G_{res} (k_1, k_2, k_3)}{k_1 k_2 k_3} =& \sum_i \frac{3 \sqrt{2 \pi} b_i^\star}{8}  \left (\frac{\sqrt{2 \epsilon_\star}}{f_i} \right ) ^{3/2} \left [ \sin \left (\frac{\phi_K+c_i}{f_i} \right )  - \frac{f_i }{\sqrt{2 \epsilon_\star}} \sum_{\ell \neq m} \frac{k_\ell}{k_m} \cos \left (\frac{ \phi_K+c_i}{f_i}\right ) \right.  \\    & - \left . \left ( \frac{f_i}{\sqrt{2 \epsilon_\star}}\right ) ^2 \frac{K(k_1^2 + k_2^2 + k_3^2)}{k_1k_2k_3} \sin \left ( \frac{\phi_K+c_i}{f_i}\right )  \right] \\[2mm]
 = &  \sum_i \frac{3 \sqrt{2 \pi} b_i^\star}{8}  \left (\frac{\sqrt{2 \epsilon_\star}}{f_i} \right ) ^{3/2} \left [ \sin  \left ( \frac{\sqrt{2 \epsilon_\star}}{f_i} \ln \frac{K}{k_\star} + \frac {c_i}{f_i} \right ) \right. + \\  & \left . \frac{f_i}{\sqrt{2 \epsilon_\star}} \sum_{\ell \neq m} \frac{k_\ell}{k_m} \cos \left ( \frac{\sqrt{2 \epsilon_\star}}{f_i} \ln \frac{K}{k_\star} + \frac {c_i}{f_i}\right ) + ... \right ].
\end{aligned}
\end{align}

As a consistency check of our calculation, we have checked that this sum of resonant bispectra satisfies the squeezed limit consistency relation~\cite{Maldacena:2002vr,Creminelli:2004yq}. The details are presented in Appendix~\ref{consistency}.

\section{Equilateral features from summed resonant non-Gaussianity}

We now want to discuss what kind of non-Gaussianities could have a sizable overlap with summed resonant non-Gaussianity. One could also turn this question around and ask what values the parameters $b_i^\star, f_i$ and $c_i$ have to take in order to generate a degeneracy. From an effective field theory point of view it is perfectly fine to choose ad hoc values for these parameters as long as they are not in conflict with observations such as the power spectrum and fulfill certain consistency conditions. For instance, the frequencies should certainly not be super Planckian, i.e. $f_i<1$, and monotonicity of the inflaton potential requires $b_i^\star < 1$. A possible stringy origin of these parameters is discussed below in Section~\ref{multiinstantoncorr}.

In~\cite{Pahud:2008ae,Flauger:2009ab}, bounds on oscillating features in the power spectrum were given for a quadratic and a linear inflaton potential respectively. The bound that was found in both works is
\begin{equation}
 b_i^\star f_i < \frac{10^{-5}}{\sqrt{2\epsilon_\star}}\,, \label{bifibound}
\end{equation}
for a single modulation in the potential,  which may be avoided if
the symmetry under time translations is collectively broken~\cite{Behbahani:2012be}.  In the following, we assume that this bound is valid for multiple modulating terms in the potential. 
For more detail on when this bound eq.~\eqref{bifibound} is applicable, see the discussion in 
Section~\ref{sec_powerspecconst}.

Let us briefly discuss the general form of the bispectrum. Due to the appearance of the delta function in eq.~\eqref{threepointGgen}, a momentum configuration is completely characterized by the absolute values of the three momenta $k_1$, $k_2$ and $k_3$. Furthermore, for a scale-invariant spectrum this reduces to two variables. Then, one usually considers the bispectrum as a function of the two rescaled momenta $x_2 = k_2/k_1$ and $x_3 = k_3/k_1$. A region that includes only inequivalent momentum configurations is given by $1-x_2\leq x_3\leq x_2$.\\
Let us switch to the variables $x_\pm = x_2 \pm x_3$, with $1 < x_+ < 2$ and $0 < x_- < 1$. Note that the resonant non-Gaussianity eq.~\eqref{sumbispectrum} is to first order in $f_i/\sqrt{2\epsilon_\star}$ only a function of $x_+$ and $k_1$ but not of $x_-$ since
\begin{equation}
 \sin \left( \frac{\sqrt{2 \epsilon_\star}}{f_i}\, \ln \frac{K}{k_\star}\right) = \sin \left( \frac{\sqrt{2 \epsilon_\star}}{f_i}\, (y + \ln k_1/k_\star ) \right) \,,\label{sinxpk1}
\end{equation}
having defined $y\equiv \ln (1+ x_+)$. Note that summed resonant non-Gaussianities are therefore not scale invariant, as is clear from in the explicit dependence on $k_1$ in eq.~\eqref{sinxpk1}. We will discuss the issue of scale dependence in Section~\ref{sec_scaledep}.

Furthermore, eq.~\eqref{sinxpk1} implies that as far as other types of non-Gaussianities are concerned, we can only expect degeneracies with the summed resonant type if they are predominantly a function of $x_+$. We will show in the next section, Section~\ref{secxm0}, that this is primarily the case for equilateral non-Gaussianity, typically arising in non-canonical models of inflation.

To measure the degree of degeneracy between different kinds of non-Gaussianities we follow~\cite{Fergusson:2008ra}. The bispectrum can always be parametrized in the form $\frac{\mathcal G (k_1, k_2, k_3)}{k_1 k_2 k_3} \sim f_{NL} S(k_1, k_2, k_3)$, with `amplitude' $f_{NL}$ and shape function $S(k_1, k_2, k_3)$.  In general, the shape refers to the dependence of $S(k_1, k_2, k_3)$ on the momentum ratios $k_2/k_1$ and $k_3/k_1$ when the overall momentum scale $K$ is fixed, while the dependence of $S$ on $K$ when the momenta $k_i$ are fixed gives the running of the bispectrum.

Now the cosine of two shapes is defined via the normalized scalar product 
\be
C(S, S')  =  \frac{F(S, S')}{\sqrt{F(S,S) F(S',S')}}, \label{cosine}
\ee
where 
\be
F(S, S') = \int_{\mathcal V} \frac{d \mathcal V}{K} S(k_1, k_2, k_3) S' (k_1, k_2, k_3) \,,\label{FSSint}
\ee
with
\begin{equation}
\frac{d \mathcal V}{K} = \frac{dk_1\,dk_2\,dk_3 }{k_1+k_2+k_3} = \frac{1}{2}k dk\,d\alpha\,d\beta\,,
\end{equation}
where we have switched to the variables $k$, $\alpha$ and $\beta$ defined according to~\cite{Fergusson:2008ra}
\begin{align}
 \begin{aligned}
&k= \frac{1}{2}(k_1+k_2+k_3)\,,\quad k_1=k(1-\beta)\,,\\
&k_2 = \frac{k}{2}(1+\alpha+\beta)\,,\quad k_3=\frac{k}{2}(1-\alpha+\beta)\,,
 \end{aligned}
\end{align}
with the integration boundaries
\begin{equation}
 \alpha \in [-1+\beta,1-\beta] \,, \qquad \beta \in [0,1] \qquad \text{and} \qquad k\in [k_{min},k_{max}]\,.
 \label{alphabetabounds}
\end{equation}

\subsection{Equilateral and local shapes for $x_- \to 0$} \label{secxm0}

Out of the many known types of non-Gaussianities,\footnote{For an overview see e.g.~\cite{Chen:2010xka}.} we discuss the following two representative types: 
The equilateral type 
\begin{equation}
 \frac{\mathcal G_{equil} (k_1, k_2, k_3)}{k_1 k_2 k_3} =  f_{NL}^{equil}\, S_{equil}(k_1,k_2,k_3)\,,\label{Gequil}
\end{equation}
with
\begin{equation}
 S_{equil}(k_1,k_2,k_3) = \frac{(k_1 + k_2 - k_3)(k_1 + k_3 - k_2)(k_3 + k_2 - k_1)}{k_1 k_2 k_3}\,,\label{Sequil}
\end{equation}
is characteristic of non-canonical inflation.

The local type is given by
\begin{equation}
 \frac{\mathcal G_{local} (k_1, k_2, k_3)}{k_1 k_2 k_3} = f_{NL}^{local}\, S_{local}(k_1,k_2,k_3)\,,
\end{equation}
with
\begin{equation}
 S_{local}(k_1,k_2,k_3) = \frac{k_1^3+k_2^3+k_3^3}{k_1 k_2 k_3}\,,
\end{equation}
and is dominant in multi-field inflation for instance.

In the limit $x_-\to 0$, the shape functions are given by
\begin{align}
 &S_{equil}^{x_-\to 0}(x_+) = \frac{4(x_+ - 1)}{x_+^2}\,,\label{Sequixp}\\
 &S_{local}^{x_-\to 0}(x_+) = \frac{4+x_+^3}{x_+^2}\,.
\end{align}
Now, one can evaluate the cosine eq.~\eqref{cosine} to find the overlap of  $S_{equil}$ and $S_{equil}^{x_-\to 0}(x_+)$:
\begin{equation}
 C(S_{equil},S_{equil}^{x_-\to 0}) = 0.93\,.
\end{equation}
Hence, the equilateral shape is well approximated by its $x_-\to 0$ limit. For the local shape the overlap is much smaller.  $S_{local}$ diverges for squeezed momentum configurations which makes it necessary to regulate the integrals that enter the cosine eq.~\eqref{cosine}. We only give an upper bound $C(S_{local},S_{local}^{x_-\to 0}) < 0.7$, which was obtained by cutting the integration boundaries eq.~\eqref{alphabetabounds} as follows:
\begin{equation}
 \beta \in [\Delta,1-\Delta]\,\qquad \text{with} \qquad \Delta = 5\cdot 10^{-5}\,.
\end{equation}

\begin{figure}[ht!]
\centerline{\includegraphics[width= 0.5\linewidth]{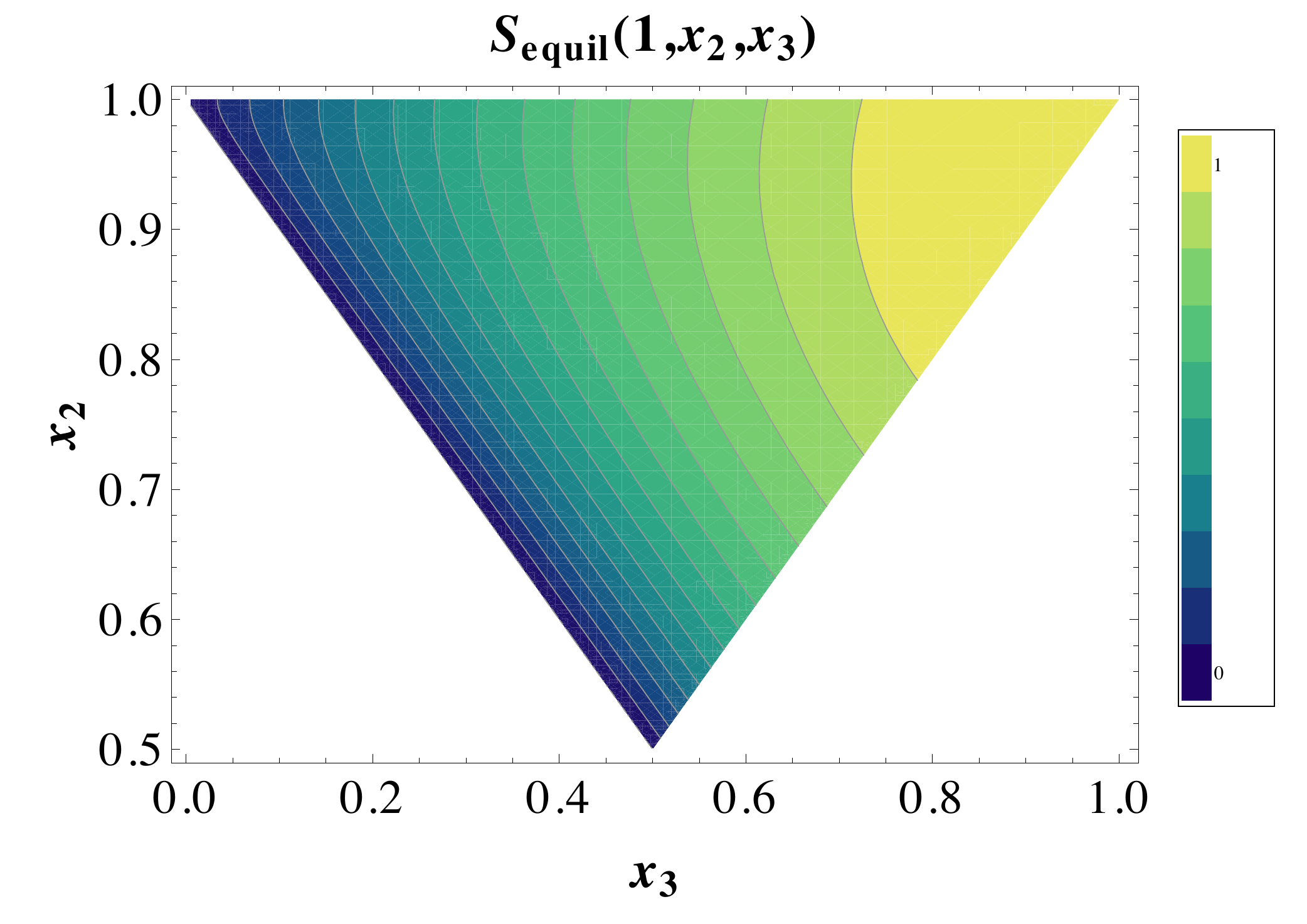}
\includegraphics[width= 0.5\linewidth]{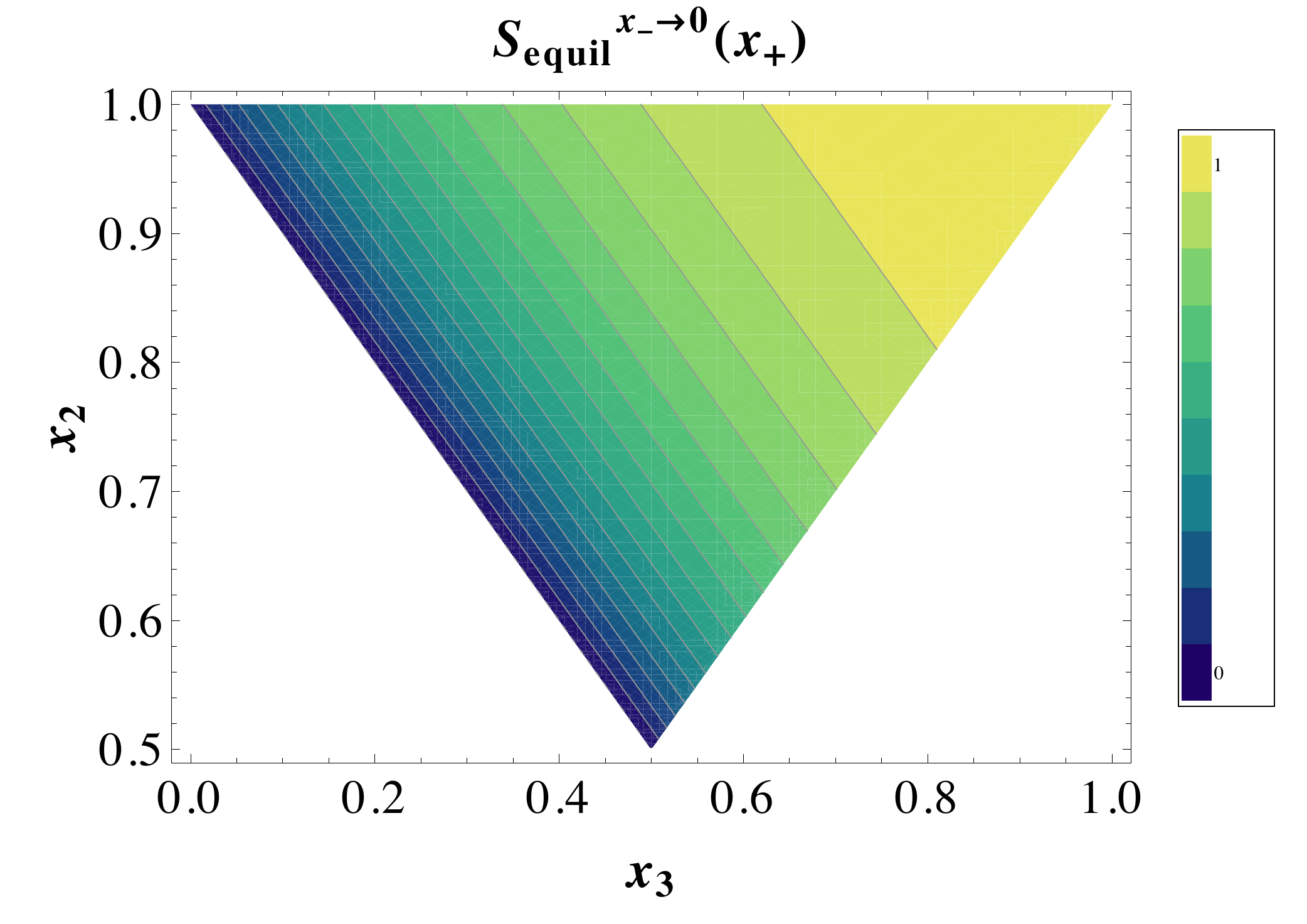}}
\centerline{\includegraphics[width= 0.5\linewidth]{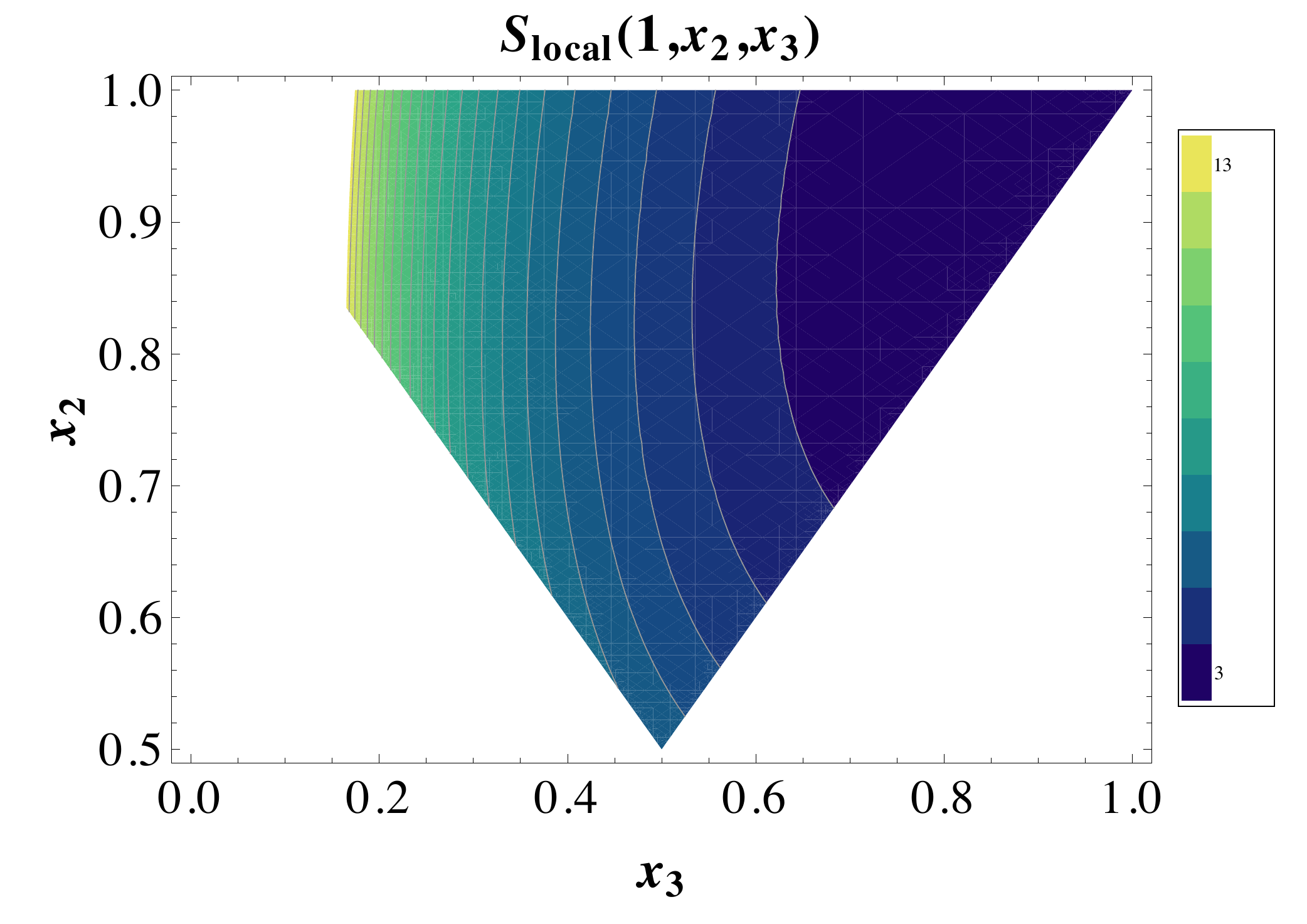}
\includegraphics[width= 0.5\linewidth]{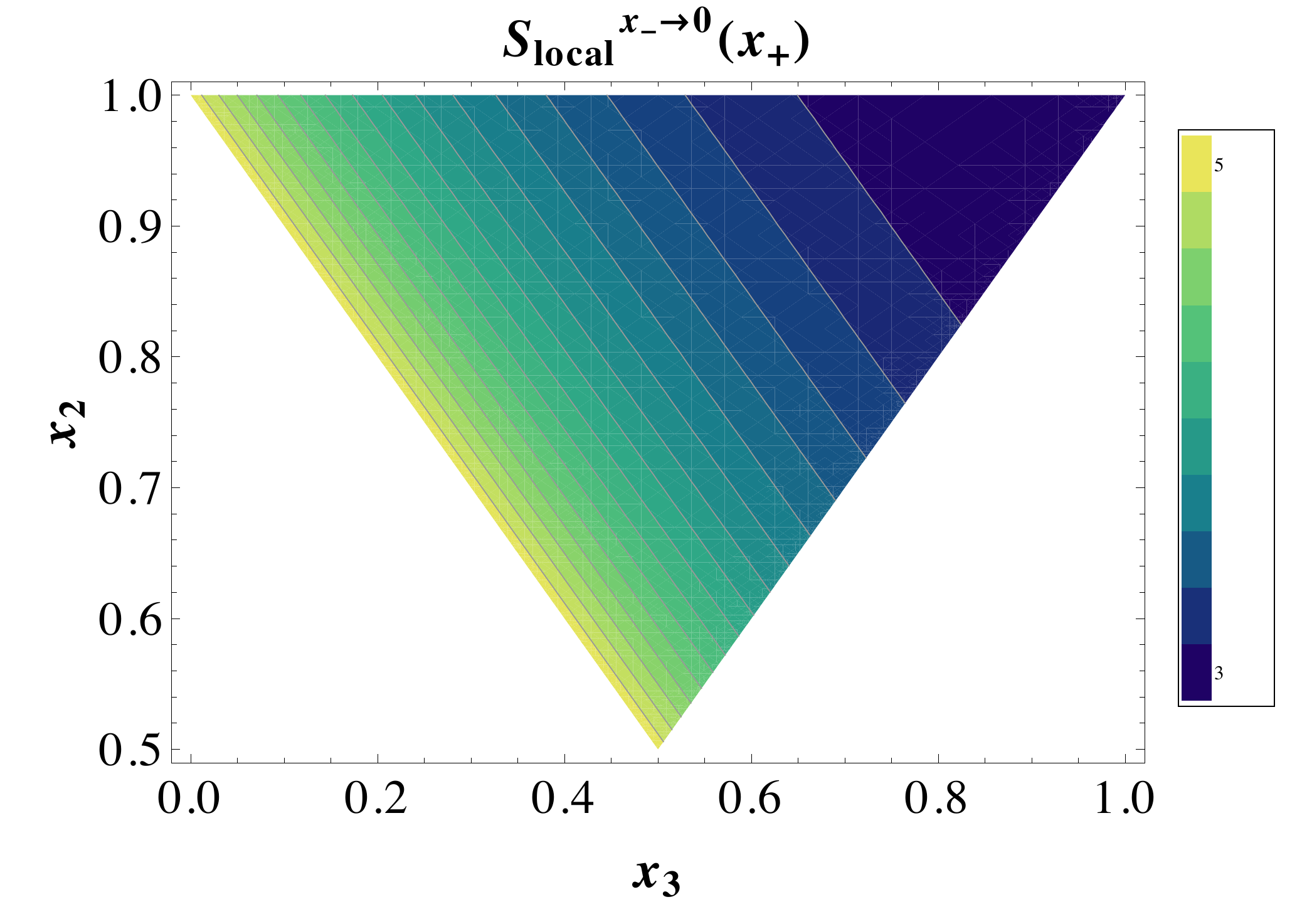}}
\caption{The equilateral and local shape functions (left) and their approximations $x_- \to 0$ (right).}
\label{equillocalcontour}
\end{figure}

The different qualitative levels of agreement of the equilateral and local shape are also visualized in Figure~\ref{equillocalcontour}. Notice that the shape function $S_{non-can}$ from non-canonical inflation shows a slightly more complicated momentum dependence than the equilateral shape function eq.~\eqref{Sequil}, see~\cite{Chen:2006nt}. We find $C(S_{non-can},S_{non-can}^{x_- \rightarrow 0})>0.93$, so the approximation by the $x_-\to 0$ limit is even better in the case of $S_{non-can}$.

\subsection{Scale invariance vs. scale dependence} \label{sec_scaledep}

Let us now discuss how well the scale-invariant equilateral shape can be approximated by a scale-dependent shape, such as summed resonant non-Gaussianity. It is obvious that the overlap cannot be made arbitrarily large; however, we will show in the following that the overlap can still be considerable.

Let $S^{per}_{equil}(y)$ be the periodic generalization of
\begin{equation}
S_{equil}^{x_-\to 0}(y) = 4 \frac{e^y - 2}{(e^y-1)^2}\,,\qquad \qquad y \in [\ln 2,\ln 3]\,,
\end{equation}
to $y\in \mathbb{R}$, i.e. $S^{per}_{equil}(y+\Delta y) = S^{per}_{equil}(y)$ with $\Delta y = \ln 3 - \ln 2$. This definition is motivated by the fact that we are going to Fourier expand on the interval $y \in [\ln 2,\ln 3]$ and this expansion will itself be periodic with period $\Delta y$.

\begin{figure}[htp]
\centerline{\includegraphics[width= 0.5\linewidth]{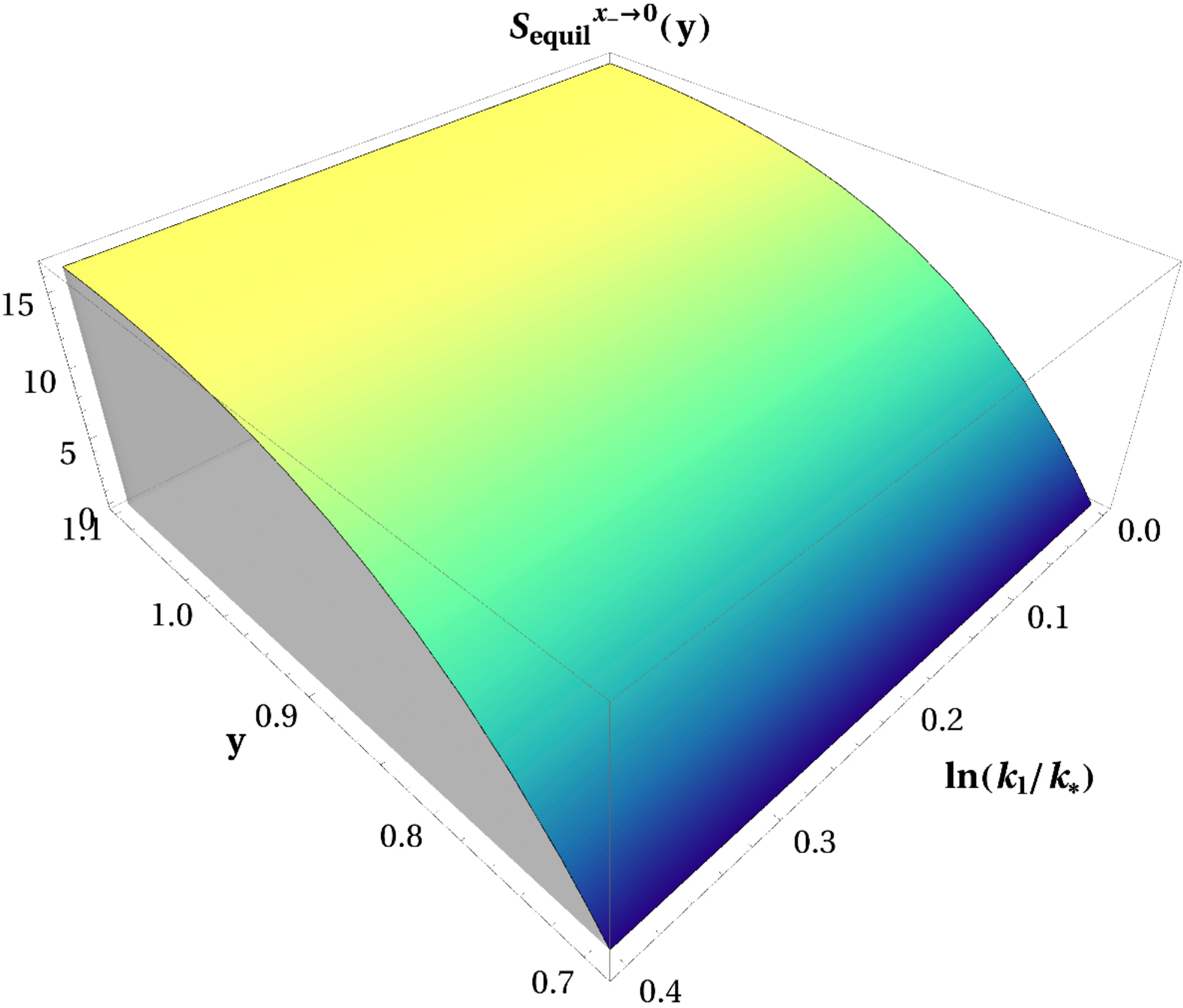}
\includegraphics[width= 0.5\linewidth]{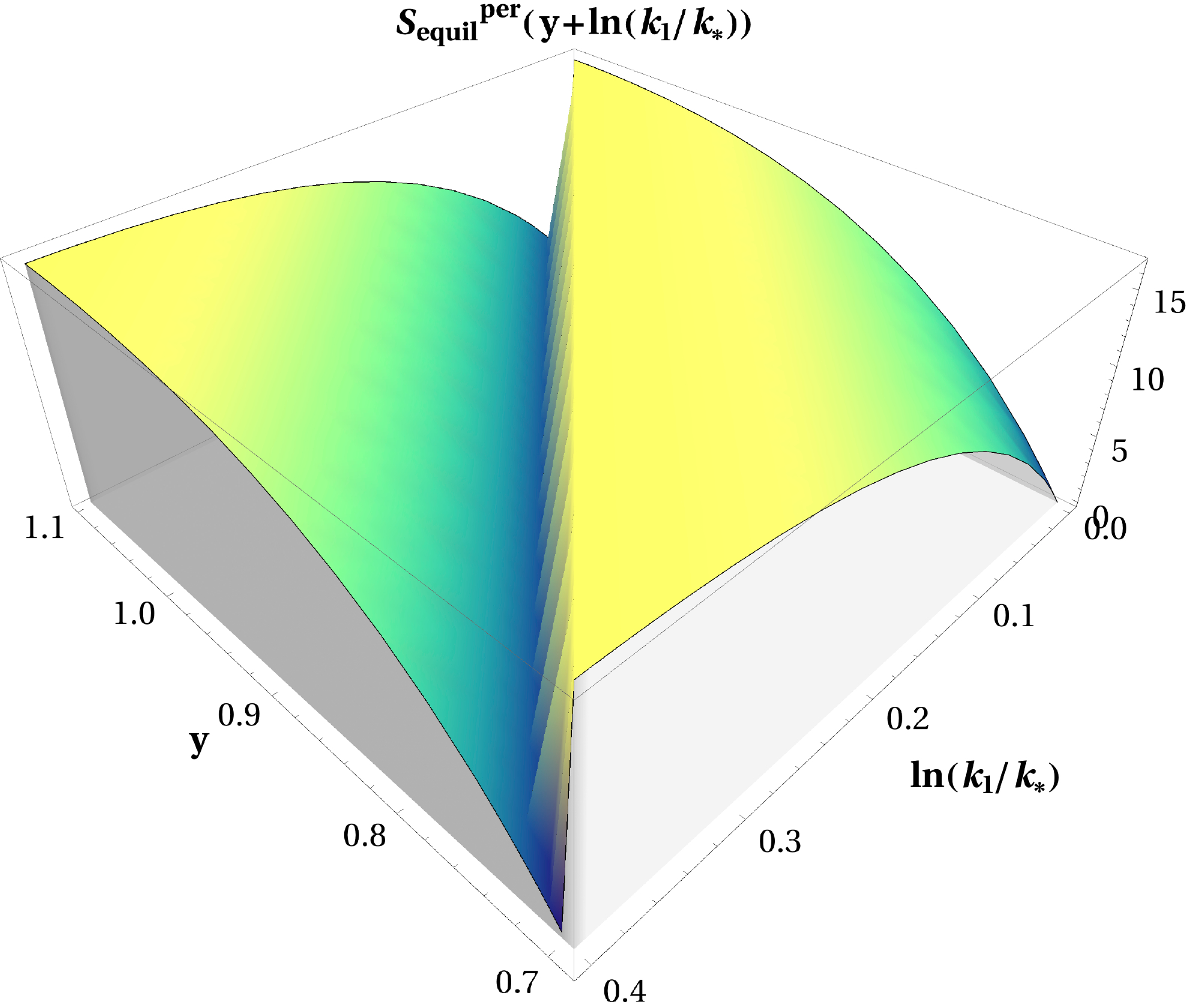}}
\caption{The scale-invariant shape $S_{equil}^{x_-\to 0}(y)$ (left) and scale-dependent shape $S^{per}_{equil}(y+\ln k_1/k_\star)$ (right) for $y \in [\ln 2,\ln 3]$ and $\ln k_1/k_\star \in [0,\Delta y]$}
\label{SperSequil}
\end{figure}

Furthermore, let us consider the scale-dependent shape $S^{per}_{equil}(y+\ln k_1/k_\star)=S^{per}_{equil}(\ln K/k_\star)$ which is a shape that can be approximated by the shape of some combination of resonant non-Gaussianities, since they have the same functional dependence, see eq.~\eqref{sinxpk1}. The different shapes in $(y,\ln k_1/k_\star)$ space are shown in Figure~\ref{SperSequil}.

The calculation of the cosine eq.~\eqref{cosine} between the shapes $S^{per}_{equil}(\ln 2k/k_\star)$ and $S_{equil}(\alpha,\beta)$ simplifies due to the reduced dependencies on the integration variables $k$, $\alpha$ and $\beta$ to
\begin{align}
\begin{aligned}
C\left(S_{equil}, S^{per}_{equil}\right) = \frac{\int d\alpha\,d\beta\, S^{equil}(\alpha,\beta)}{\left( 2 \int d\alpha\,d\beta\, {S^{equil}}^2(\alpha,\beta)\right)^{1/2}} \frac{2\int_{k_{min}}^{k_{max}} dk \, k \, S^{per}_{equil}(\ln 2k/k_\star)}{\left([k_{max}^2-k_{min}^2]\int_{k_{min}}^{k_{max}} dk \, k \, {S^{per}_{equil}}^2(\ln 2k/k_\star) \right)^{1/2}}
\end{aligned} \label{Cosperequilform}
\end{align}
We find that eq.~\eqref{Cosperequilform} is to very good approximation independent of the values of $k_{min}$ and $k_{max}$, if $k_{min}\ll k_{\star}$ and $k_{max} \gg k_{\star}$. This is due to the periodicity of $S^{per}_{equil}(\ln 2k/k_\star)$, yielding the second fraction on the RHS of eq.~\eqref{Cosperequilform} essentially independent of these values. For $k_{max}=10^4 k_\star$ and $k_{min}=10^{-4} k_\star$, we find
\begin{equation}
 C\left(S_{equil}, S^{per}_{equil}\right) = 0.83\,.
\end{equation}

Let us also mention that the overlap would be greater than $83 \%$, if we had considered not the equilateral shape function eq.~\eqref{Sequil}, but the exact shape function~\cite{Chen:2006nt} that arises in non-canonical single field models of inflation. This is due to the fact that the overlap between $S^{x_- \to 0}$ and $S$ is larger than for the equilateral shape function eq.~\eqref{Sequil}, as discussed at the end of Section~\ref{secxm0}.

In the light of data analysis, let us discuss potential ways to break the degeneracy between $S_{equil}$ and $S^{per}_{equil}$. There is a certain overlap between the local and equilateral shape function~\cite{Fergusson:2008ra}. Note that the local shape diverges in the squeezed limits $k_i \to 0$ and hence the integral eq.~\eqref{FSSint} requires regularization which also implies that the final result for the cosine is regularization dependent. Here we pick the cut-off regularization
\begin{equation}
 S_{local}^{reg}=\begin{cases} S_{local} &\mbox{if } k_i / \sum_{j\neq i}^3 k_j \geq 0.01 \\
0 & \mbox{else}\,. \end{cases} \label{reglocal}
\end{equation}
The correlation between the equilateral respectively periodic equilateral shape and the local shape are
\begin{equation}
 C\left(S_{equil}, S_{local}^{reg}\right) = 0.47\,,\qquad \qquad C\left(S_{equil}^{per}, S_{local}^{reg}\right) = 0.68\,.
\end{equation}
Assuming the measured CMB bispectrum is primarily of the equilateral or periodic equilateral type one can use different shape templates in data fitting to find out the true shape of the bispectrum. For instance, if the CMB bispectrum shape is equilateral, fitting with the equilateral, periodic equilateral and local shape will yield a $100\%$, $83\%$ and $47\%$ correlation while if it is of the periodic equilateral type a $83\%$, $100\%$ and $68\%$ correlation will result. Hence, using different templates that have an overlap with the equilateral shape, e.g. the local template, could potentially help to distinguish between the significantly overlapping equilateral and periodic equilateral shapes in the CMB data. Furthermore, as we will see in the following Section~\ref{ref_fourieran}, oscillatory features remain in the squeezed limit even if the periodic equilateral shape is synthesized in a Fourier analysis. An observation of such oscillatory features in the CMB bispectrum, see~\cite{Fergusson:2010dm}, would 
also break the degeneracy.

\subsection{Fourier analysis}    \label{ref_fourieran}

Having found that the equilateral shape function has a non-negligible degeneracy with the one dimensional function $S^{per}_{equil}(y+\ln k_1/k_\star)$, the question of degeneracy with summed resonant non-Gaussianities can be expressed in terms of a Fourier analysis. Let us set $k_1=k_\star$ for conciseness in the remainder of this section; however the following analysis is valid for all values of $k_1$.

Let us define
\begin{equation}
 B_i = \frac{3\sqrt{2\pi}b^\star_i}{8} \, \left( \frac{\sqrt{2\epsilon_\star}}{f_i} \right)^{3/2}\,\quad \text{and} \quad F_i = \frac{\sqrt{2\epsilon_\star}}{f_i}\,. \label{BiFi}
\end{equation}
Then, summed resonant non-Gaussianity, eq.~\eqref{sumbispectrum}, can, to first order in $f_i / \sqrt{2 \epsilon_\star}$, be written in the form\footnote{Due to the suppression by $f_i / \sqrt{2\epsilon_\star} \ll 1$, the second order contribution to $\mathcal G_{res}/(k_1 k_2 k_3)$  induces a non-negligible correction only in the squeezed momentum configurations which we {\it discuss at the end of this section}.}
\begin{align}
 \begin{aligned}
  \frac{\mathcal G_{res}}{k_1 k_2 k_3} (y) &= \sum_{i=1}^{2N} B_i \sin(F_i y + C_i) =\sum_{i=1}^{N} B_i \cos(F_i y) + B_{N+i} \sin(F_i y)
\label{GresFourier}
 \end{aligned}
\end{align}
where in the last equality of eq.~\eqref{GresFourier}, we have chosen the phases $C_i$ such that there appears a sine and a cosine for each frequency $F_i$.

The Fourier expansion of a generic function $f(y)$ on the interval $y\in [a,a+b]$ is given as
\begin{equation}
 f(y) \simeq \frac{u_0}{2} + \sum_i^N \left[u_i \cos \left(\frac{2 \pi i}{b} y \right) + u_{N+i} \sin \left(\frac{2 \pi i}{b} y \right) \right]\,,\label{FourierGen}
\end{equation}
with
\begin{align}
 \begin{aligned}
  u_i &= \frac{2}{b} \int_a^{a+b} f(y) \cos \left(\frac{2 \pi i}{b} y \right) dy\quad \text{for} \,\, i\geq 0\,,\\
  u_{N+i} &= \frac{2}{b} \int_a^{a+b} f(y) \sin \left(\frac{2 \pi i}{b} y \right) dy\quad \text{for} \,\, i\geq 1\,.
 \label{fouriercoeffgen}
 \end{aligned}
\end{align}

Note that eq.~\eqref{GresFourier} is of the Fourier form, eq.~\eqref{FourierGen}, if the constant term
\begin{equation}
 u_0 = \frac{2}{b} \int_a^{a+b} f(y) dy\,,\label{u0}
\end{equation}
is zero. Since $S^{per}_{equil}(y)$ is monotonically increasing and has a zero at $\ln 2$, the numbers $a$ and $b$ can always be chosen such that $u_0=0$. Note that we can always redefine $S^{per}_{equil}(y)$ on the (non-physical) interval $[\ln 2 - \xi,\ln 2]$ for $\xi\ll1$ such that the integral eq.~\eqref{u0} has a large negative contribution on this interval, i.e.
\begin{equation}
 u_0 \sim \int_a^{a+b} f(y) dy = \int_{\ln 2 - \xi}^{\ln 2} f(y) dy + \int_{\ln 2}^{\ln 3} f(y) dy\,,\label{whyu0zero}
\end{equation}
where the first term on the RHS of eq.~\eqref{whyu0zero} is negative and has the same absolute value as the positive second term such that $u_0 = 0$. For instance one could use $f(y)=-|f_0|$ with $|f_0|\gg1$ for $y \in [\ln 2 - \xi,\ln 2]$. Hence, $b=\ln 3 -\ln 2$.

\begin{figure}[t]
\centerline{\includegraphics[width= 0.5\linewidth]{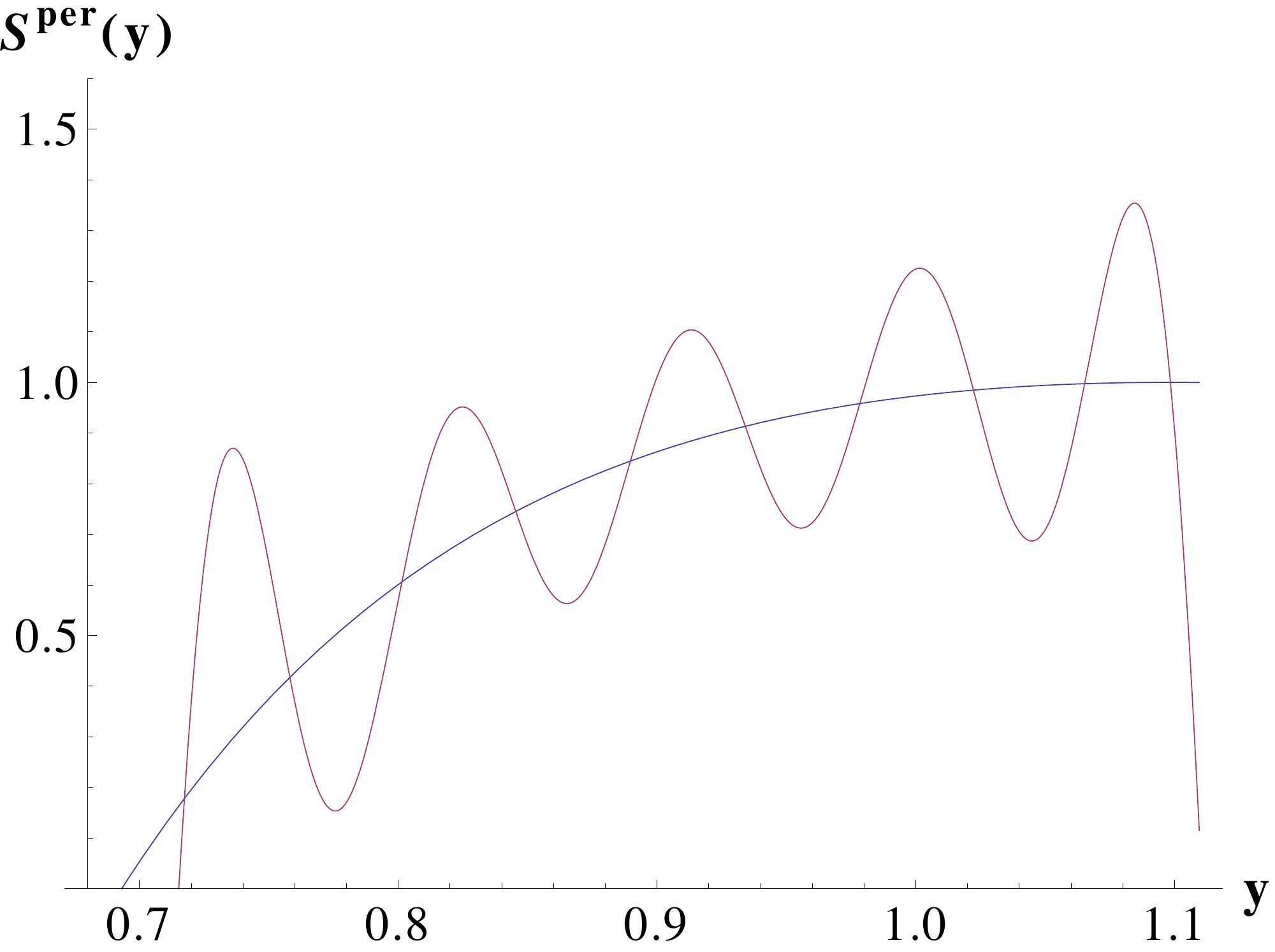}
\includegraphics[width= 0.5\linewidth]{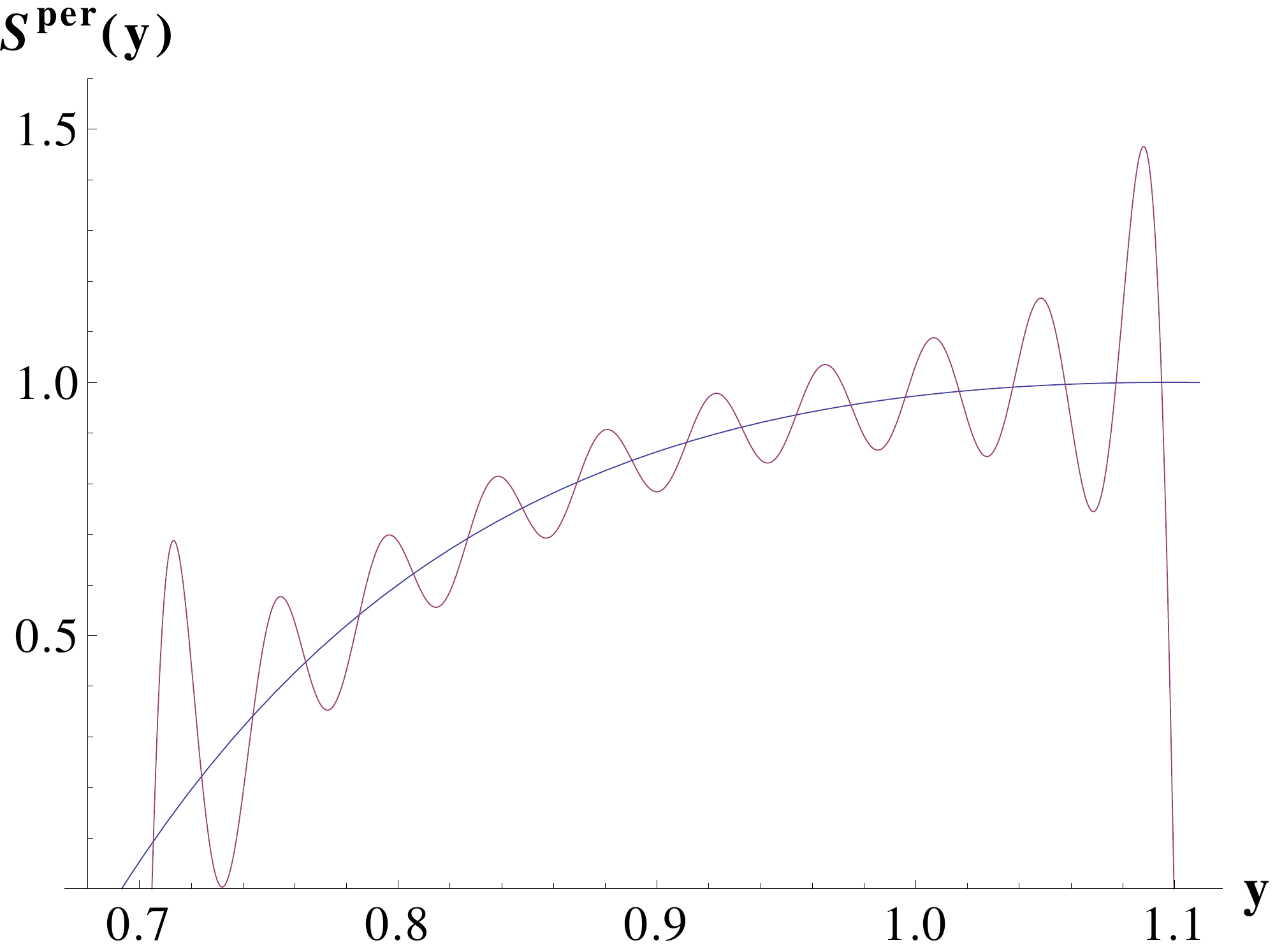}}
\caption{Fourier expansion of $S^{per}_{equil}(y)$ for $N=5$ (left) and $N=10$ (right).}
\label{onedimfourier}
\end{figure}

\begin{figure}[ht!]
\centerline{\includegraphics[width= 0.5\linewidth]{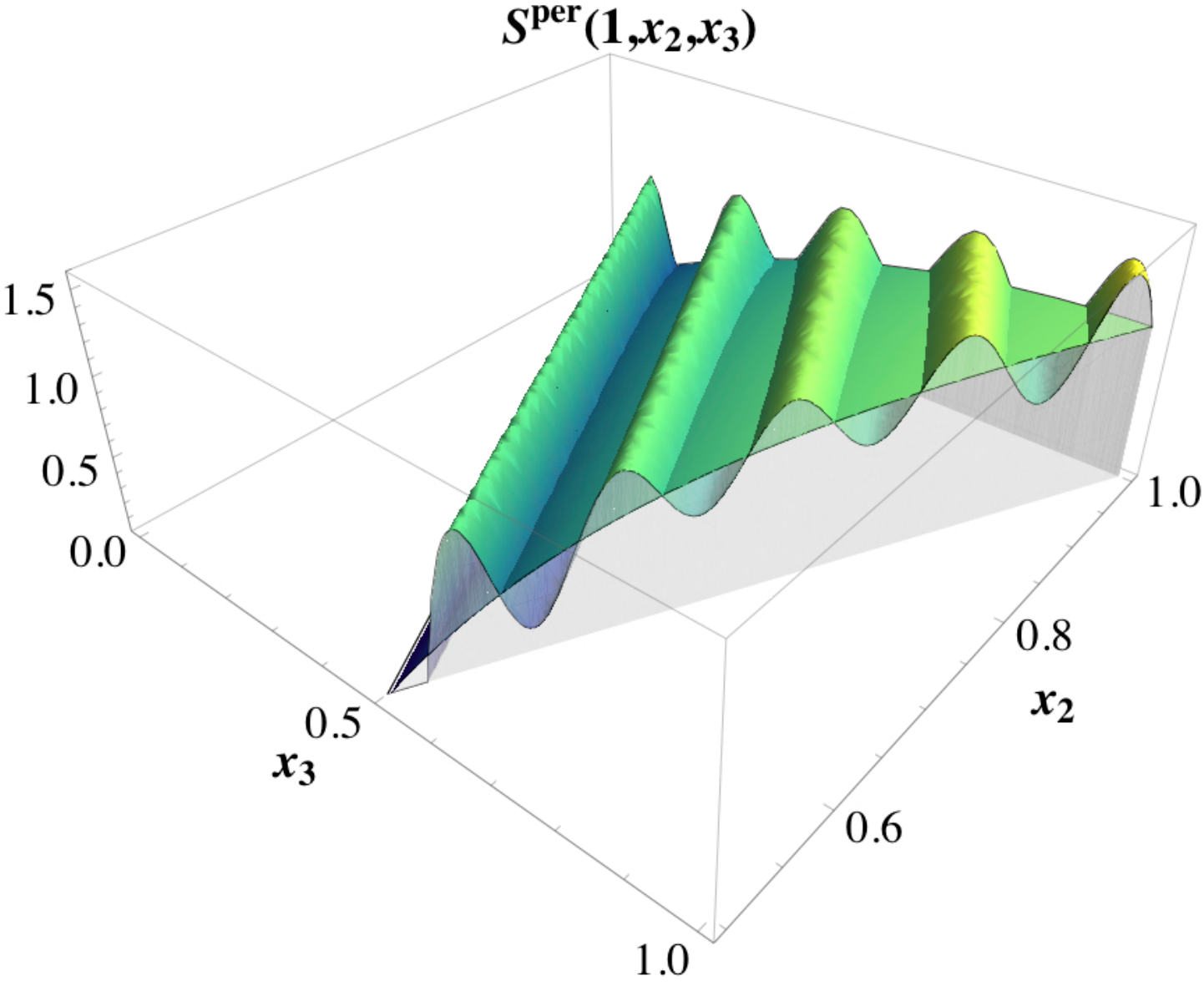}
\includegraphics[width= 0.5\linewidth]{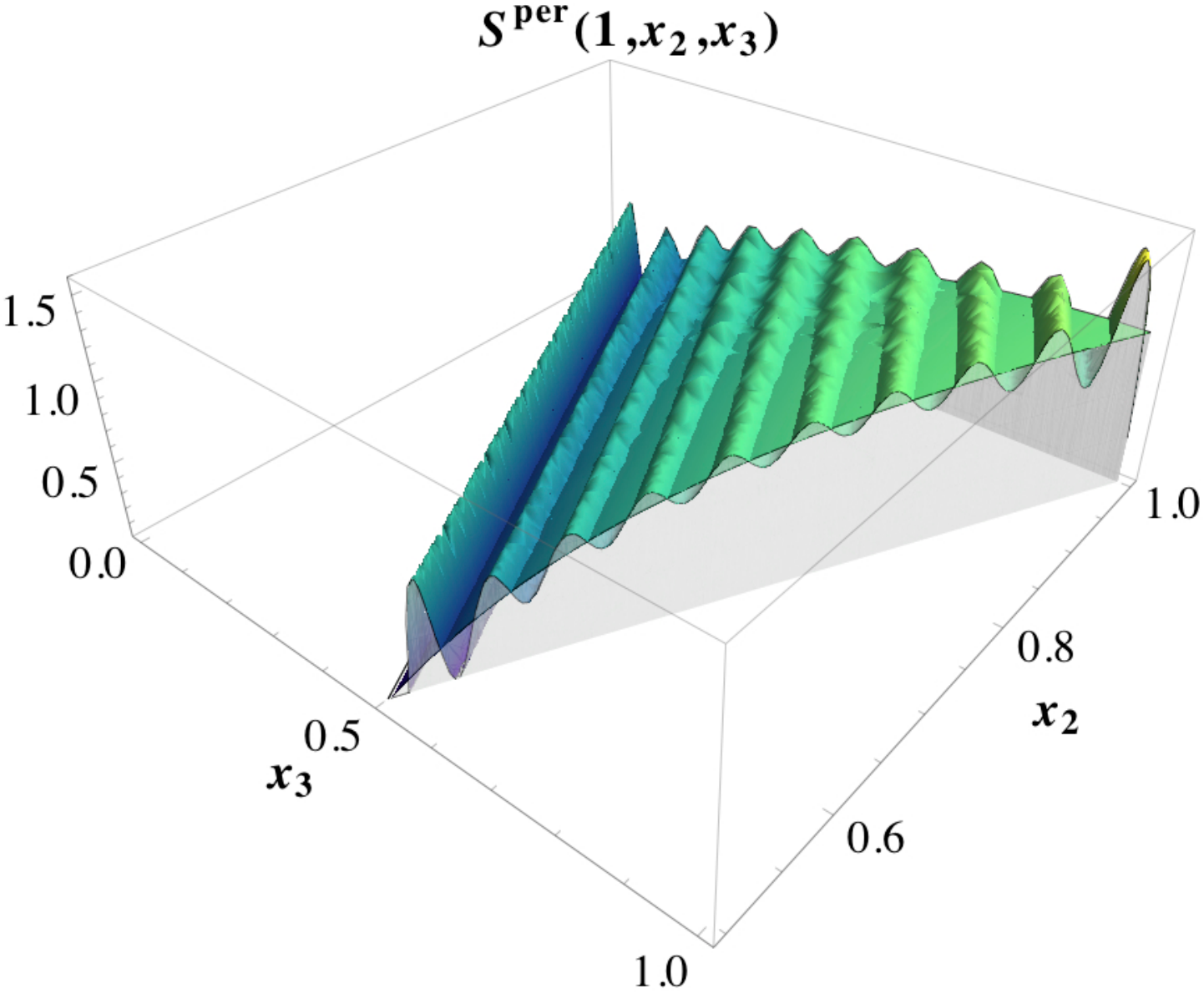}}
\caption{Fourier expansion of $S^{per}_{equil}(y)$ for $N=5$ (left) and $N=10$ (right). We have set $k_1=k_\star$.}
\label{twodimfourier}
\end{figure}

In Figures~\ref{onedimfourier} and~\ref{twodimfourier}, we show the Fourier expansion of $S^{per}_{equil}(y)$ for $N=5$ and $N=10$. The coefficients $u_i$ are given in table~\ref{fouriercoeff} for $N=5$.  Note that the sum eq.~\eqref{FourierGen} is generically dominated by the low frequency terms $i\gtrsim 1$ since they do not oscillate much on the domain $[\ln 2-\xi, \ln 3]$ as can be seen from the hierarchy in the coefficients $u_i$ in table~\ref{fouriercoeff}. The higher order coefficients $i\lesssim N$ approximate the periodic equilateral shape to higher precision.
\begin{table}[t!]
\centering
  \begin{tabular}{|c|c|c|c|c|c|c|c|c|c|}
\hline
  $u_1$ & $u_2$ & $u_3$ & $u_4$ & $u_5$ & $u_6$ & $u_7$ & $u_8$ & $u_9$ & $u_{10}$\\
  \hline
  0.41 & 0.01 & -0.17 & 0.04 & 0.09 & -0.20 & 0.24 & -0.03 & -0.12 & 0.05 \\
\hline  
\end{tabular}
  \caption{Fourier expansion coefficients $u_i$, eq.~\eqref{fouriercoeffgen}, of $S^{per}_{equil}(y)$ for $N=5$.}
  \label{fouriercoeff}
\end{table}

Let us discuss how the Fourier expansion parameters $a$, $b$ and $u_i$ of eq.~\eqref{fouriercoeffgen} relate to the parameters $B_i$ and $F_i$ of eq.'s~\eqref{BiFi} and~\eqref{GresFourier}. Clearly, if we want to impose
\begin{equation}
 \mathcal G^{per}_{equil}(y) \simeq \mathcal G_{res} (y)\,,
\end{equation}
we have to set
\begin{equation}
 B_i = f_{NL}^{equil}\, u_i \leq 0.41\,f_{NL}^{equil} \,\qquad \text{and} \qquad F_i = \frac{2 \pi i}{b} \simeq 15.5\,i\,,\label{BiuiFib}
\end{equation}
where we have used that the maximal $u_i$ is $u_1 = 0.41$ for $N=5$. In the derivation of the bispectrum in Section~\ref{thebispectrum} we assumed $f_i \ll \sqrt{2\epsilon_\star}$, i.e. $F_i \gg 1$. We see from eq.~\eqref{BiuiFib} that this constraint is satisfied with growing $i$. Now, solving for $b^\star_i$ and $f_i$ in eq.~\eqref{BiFi} yields
\begin{equation}
 b^\star_i = \frac{8}{3\sqrt{2 \pi}}\,\frac{f_{NL}^{equil}\,u_i}{F_i^{3/2}} \leq \frac{f_{NL}^{equil}}{f_{NL}^{max}}\,\qquad \text{and} \qquad f_i = \frac{\sqrt{2\epsilon_\star}}{F_i} = 0.06 \frac{\sqrt{2\epsilon_\star}}{i}\,,\label{bififNL}
\end{equation}
where we have once again used $\max \{u_i\} = 0.41$ and $f_{NL}^{max}\equiv 140$.

There are two things we can conclude from eq.~\eqref{bififNL}. First of all, frequencies are necessarily sub-Planckian as is required for consistency. Second, we see that for $f_{NL}^{equil} > f_{NL}^{max}$ the monotonicity condition $b^\star_i < 1$ is violated. Hence $f_{NL}^{max}\equiv 140$ is the maximal equilateral non-Gaussianity that could be matched by summed resonant non-Gaussianity if there were no constraints from the power spectrum. 

To conclude this section let us discuss the role of the second order corrections that become sizable in the very squeezed limit to the resonant bispectrum given in eq.~\eqref{sumbispectrum}. Since these corrections come with an inverse power of the Fourier frequencies $F_i = \sqrt{2\epsilon_\star}/f_i$ and generically $u_1 \gg u_i$ for $i\lesssim N$ the corrections are dominated by the low frequency terms $i\gtrsim1$. Hence, assuming that $S^{per}_{equil}$ can be approximated to arbitrary precision in a Fourier analysis we define the shape function
\begin{equation}
 S^{per-corr}_{equil} \equiv S^{per}_{equil} + \frac{1}{F_1} \,\sum_{l\neq m} \frac{k_l}{k_m} \cos \left(F_1 \ln \frac{K}{k_\star} \right)\,,
\end{equation}
with $F_1 \simeq 15.5$ to estimate the effect of these corrections. To calculate the cosine the divergency of this shape function in the squeezed limit $k_i \to 0$ is regulated by the same cutoff that was used in the case of the local shape function eq.~\eqref{reglocal}. We find
\begin{equation}
 C\left(S_{equil}, S^{per-corr}_{equil}\right) = 0.77\,,
\end{equation}
compared to $83\%$ correlation between $S_{equil}$ and $S^{per}_{equil}$. As expected the second order terms further break the degeneracy with the equilateral shape in the very squeezed limit. However, the effect is moderate.

\subsection{Constraints from the power spectrum} \label{sec_powerspecconst}

Let us discuss the constraint on the product $b^\star_i f_i$ from the power spectrum, i.e. eq.~\eqref{bifibound}. As discussed in the Introduction, non-observation of oscillating contributions to the 2-point function can place tight constraints on the maximum amount of resonantly generated non-Gaussianity with equilateral characteristics. However, the power spectrum constraint can be evaded when the time translation symmetry (which gives rise to scale invariance) is collectively broken as in \cite{Behbahani:2012be}. Here, we discuss the bound on $f_{NL}$ arising from the constraints on oscillations in the power spectrum, and also the mechanism of collective symmetry breaking whereby this constraint may be avoided.

First, using eq.~\eqref{bififNL}, we find
\begin{equation}
 b^\star_i f_i < \frac{8}{3\sqrt{2 \pi}}\,\frac{\sqrt{2\epsilon_\star}\, f_{NL}^{equil}\,u_i}{F_i^{5/2}} < 5\cdot 10^{-4} f_{NL}^{equil}\, \sqrt{2\epsilon_\star}\,,
\end{equation}
where the product on the far right is the maximum value of $ b^\star_i f_i $. Since the maximum value of $b^\star_i f_i$ has to be smaller than the upper bound given in eq.~\eqref{bifibound} this implies
\begin{equation}
 \epsilon_\star\,f_{NL}^{equil}   < 10^{-2}\,. \label{epsstarfnlineq}
\end{equation}

Given the values for $b^\star_i$ and $f_i$ obtained for the Fourier expansion, we can check the behaviour of eq.~\eqref{deltans}. We find that to leading order it behaves like a single oscillation with $f_i$ given by the lowest frequency $f_1$. This means that the two-point function bounds found for the singly modulated potential will still apply in our case, assuming no other U(1)s are present, as in \cite{Behbahani:2012be}. 

Note, that eq.~\eqref{bifibound} was derived for large-field models (linear axion monodromy inflation~\cite{Flauger:2009ab} and a quadratic potential~\cite{Pahud:2008ae}), and thus should not have validity for, say, $\epsilon_\star\lesssim 10^{-3}$. Therefore, from this result we can argue at most for a resonantly generated $f_{NL}^{equil}\lesssim {\cal O}(1)$.

This contraint from the 2-point function bound can be understood from the fact that the resonant $N$-point functions in inflationary models with periodically broken shift symmetry display a hierarchical suppression with increasing $N$~\cite{Behbahani:2011it}. However, if the mechanism of shift symmetry breaking is collective (such that scale invariance is protected by several independent symmetries), this hierarchy is no longer present. Any scale-dependent correlation function must depend on all the couplings required to break the symmetry, so that it is possible to have a nearly scale-invariant power spectrum and large non-Gaussianity (scale dependence in the bispectrum), as was shown for resonantly generated non-Gaussianity in \cite{Behbahani:2012be}. In this case, for instance when an extra global U(1) symmetry is present, the constraints discussed above can be avoided, and a sizable equilateral type $f_{NL}$ (up to 140) could be produced via the summation discussed here, without implying large 
oscillations in the power spectrum.

The limit eq.~\eqref{epsstarfnlineq} was derived for a large-field model by imposing the limit which the 2-point function data places on this model in eq.~\eqref{bifibound}. As such, we cannot apply our bound to small-field models. However, two further comments are in order. Firstly, in the absence of a collective breaking of the shift symmetry by, say, additional global U(1) symmetries~\cite{Behbahani:2012be}, the structure of the resonant N-point functions places a tight constraint $f_{NL}\sim {\cal O}(5)$~\cite{Behbahani:2011it}, regardless of the field range during inflation. For a parametrically small-field model ($\Delta\phi_{60\,e-folds}\ll 1$), we would have to get multiple instanton contributions with axion decay constants with even parametrically smaller values in order to produce many wiggles in the potential within the 60 e-fold field range. As the axion decay constants are typically given by an ${\cal O}(1)$ inverse power of the size of the extra dimensions, they are difficult to get very 
small. Therefore, we expect parametrically small-field models to appear generically with few or no significant oscillatory contributions within their 60 e-folds field range. If this is true, then small-field models might be generically expected to produce no oscillations in the 2-point function and no resonant non-Gaussianity at all.

\section{Possible avenues for embedding: Axion monodromy inflation} \label{multiinstantoncorr}
We have seen that a potential modulated by a sum of small oscillating terms  as given in eq.~\eqref{sum_pot} : 
\be
\label{sumpot2}
V(\phi) = V_0 (\phi) + \sum_i A_i  \cos \left (\frac{\phi + c_i}{f_i} \right),
\ee
will lead to a bispectrum given by a sum of resonant bispectrum terms eq.~\eqref{sumbispectrum}. This follows as long as $f_i \ll \sqrt{2 \epsilon_\star}$ for each $i$. Furthermore, a Fourier series of  these resonant bispectra leads to a summed resonant bispectrum which closely approximates the equilateral bispectrum. Taking this series amounts to a choice of parameters $\{ f_i, c_i\}$, consistent with the bounds above, for a suitable number of terms in the bispectrum or equivalently in the modulated potential eq.~\eqref{sum_pot}.

 It remains to ask how such a sum of sinusoidal corrections to the potential could arise. In this section we argue for one possible source for these corrections in the context of axion monodromy inflation~\cite{McAllister:2008hb, Flauger:2009ab}, where the resonant non-Gaussianity shape is observed~\cite{Flauger:2010ja}. The oscillatory correction considered in these papers arises from nonperturbative effects such as instantons, and is quite general in large field models. We show that these effects can also give rise to a sum of periodic potentials such as that in eq.~\eqref{sumpot2} and studied here. 
 
 \subsection{Axion monodromy inflation}
 
Up to this point we have taken the potential eq.~\eqref{sumpot2} as the starting point, in keeping with a low-energy effective field theory point of view of inflation in which we are agnostic as to the UV origins of the theory.  One possible avenue for a realization of multiple resonant non-Gaussianity is given by axion monodromy inflation \cite{McAllister:2008hb, Flauger:2009ab}. Axion monodromy inflation is a string theoretic realization of large field inflation, where the approximate shift symmetry of the axion protects the required flatness of the potential from higher-dimensional operators which become relevant when the field range is super-Planckian.  Axions in string theory arise either as the Hodge duals of 2-form fields in 4 dimensions or as zero modes of p-forms on the compact manifold \cite{Svrcek:2006yi}, and inherit their periodicity from these fields. Normalization of the axion kinetic term results in a coupling to other fields inversely proportional to $f_a$, the axion decay constant. In 
general 
string theoretic constructions this decay constant 
is larger than the range allowed by cosmological bounds (to avoid excessive production of axionic dark matter which could overclose the universe) \cite{Svrcek:2006yi}. However, despite this, the field range spanned within a single period is generally sub-Planckian in string theory \cite{Banks:2003sx}. 

This limitation (for the purposes of large field inflation) is circumvented in axion monodromy inflation, where a large field range for the inflaton axion field $c = 2 \pi \int_{\Sigma} C_2$ is possible for instance when the two-cycle $\Sigma$ on which it is supported is wrapped by a space-filling 5-brane which  breaks the shift symmetry $ c \rightarrow c + 2 \pi$ in the potential, and gives rise to monodromy in the potential energy: the potential is no longer a periodic function of the axion, and increases without bound as the axion VEV increases. 
This follows from the DBI action, which gives a potential 
\begin{equation}
 V_0(c) \,\, \sim \, \, \sqrt{\ell^4 + c^2} \,,
\end{equation}
where $\ell \sqrt{\alpha '}$ is the size of the wrapped cycle. Generalizations of this effect arise generically in the presence of fluxes.

The initial axion VEV is large, leading to a linear unmodulated potential $V_0(\phi)$, which means that  $\frac{1}{\sqrt{2 \epsilon_\star}} = \phi_\star $, the value of the inflaton at horizon exit, and $b^\star = b$ is independent of $\phi_\star$. Inflation ends when at small values of the axion VEV, $V_0(\phi)$ is no longer linear, the axion oscillates about its minimum, and the inflationary energy is diverted into any string modes coupled to the axion.

Many of the consistency conditions (such as e.g.~the absence of large back reaction from the axion-induced 3-brane charge building up on the 5-branes) for axion monodromy inflation were already discussed in~\cite{McAllister:2008hb, Flauger:2009ab}. Some remaining questions may consist of e.g. finding an explicit realization on a concrete compact 3-fold in type IIB (despite the generality of the mechanism which in the context of fluxes~\cite{Dong:2010in} should lead to many explicit avenues for construction), or details of the 3-brane charge back reaction and the mass scale of some of the KK modes in particular setups. For a more detailed discussion of some of these and related points see e.g.~\cite{Flauger:2009ab,Barnaby:2011qe}.

 \subsection{Periodic corrections to the potential} 
In \cite{McAllister:2008hb, Flauger:2009ab}, working in an $O3/O7$ orientifold of Type IIB,  the axion found to be suitable for this model of inflation was that arising from the zero mode of the RR two-form $C_2$ on a 2-cycle $\Sigma$ in the internal manifold, denoted  $c = 2 \pi \int_{\Sigma} C_2$. The cohomology groups $H^{(p,q)}$ split into two eigenstates under the action of the orientifold in a flux compactification \cite{Giddings:2001yu} (see also \cite{Dasgupta:1999ss})
\begin{equation}
H^{(p,q)} \,\, = H_+^{(p,q)} \oplus H_{-}^{(p,q)}\,,
\end{equation}
where the $\pm$ subscripts refer to even/odd behaviour under the orientifold action. Because $C_2$ is odd under the orientifold projection, only the mode coming from wrapping a 2-cycle which is also odd under the orientifold will survive, i.e. we take $\Sigma \in H_2^{-}$.  Such an odd cycle can be written as $v^- = v^1 - v^2$, where $v^{1}$ and $v^2$ are two-cycles in the CY which are mapped into each other by the orientifold action. The $+$ combination is then obviously an even two-cycle. 

Generally, one expects non-perturbative effects to lead to small periodic corrections to the potential for the axion, leading to the modulated potential which gives rise to resonant-type non-Gaussianity \cite{Flauger:2009ab, Flauger:2010ja}. These corrections arise from instanton corrections to the action, specifically worldsheet instanton contributions to the K\"ahler potential.  Instantons arise when the (p+1)-dimensional world volume of a Euclidean p-brane is wrapped over a (p+1) cycle in the internal manifold. Their corrections are exponentially suppressed by the size of the wrapped cycles. Space-time instantons, which correct the superpotential, are localized in $R^4$, while worldsheet instantons, which correct the K\"ahler potential, are localized in 2 dimensions. These are due to Euclidean D1-branes  and their $SL(2, \mathbb Z)$ images (i.e. F- and D-strings as well as their bound states, $(p,q)$ strings)  wrapping 2-cycles $v^+$  threaded by the axionic $C_2$ forms. In \cite{Flauger:2009ab} this 
correction 
was taken to be of the form
\bea
\label{onetermcorrection}{\mathcal K}& = &  - 2 \log [{\mathcal V}_E + e^{- 2 \pi v^+/\sqrt{g_s}} \cos (c)]\,,
\eea
for the $c$ axion, where $\mathcal{V}_E$ is the Einstein-frame volume of the CY manifold, and $v^+$ is the even cycle dual to that wrapped by $C_2$ and on which the instanton is supported. $\tau$ is the axion-dilaton $\tau = C_0 + i e^{- \phi}$  and the $G^a$ are complex scalar fields composed of the RR and NSNS axionic fields $c$ and $b$ as \cite{Grimm:2004uq}
\be
G^a  =  c^a - \tau b^a,
\ee
where $a$ runs over the number of axions.

The expression in eq.~\eqref{onetermcorrection} captures the important sinusoidal features of the instanton corrections to the K\"ahler potential. The detailed form of one such series of leading (in the sense of exponentially unsuppressed) corrections in the $b$-axions was found in \cite{Grimm:2007xm}. It is given by 

\be
g (\tau, \bar \tau, G^a,  \bar G^a)   =    - \frac{1}{4}\sum_\beta n_{\beta} \sum_{(m,n)} \frac{(\tau - \bar \tau)^{3/2}}{|n + m \tau|^3} \cos (( n + \tau m) \frac{K_a(G^a - \bar G^a)}{\tau - \bar \tau} - m k_a G^a), 
\ee
where $\beta$ are curves in the negative eigenspace $H_2^{-}$ spanned by the basis $\{ \omega^a \}$, the $n_\beta$ are the Gopakumar-Vafa invariants for these curves and   $K_a = \int_{\beta} \omega_a$. The sum over $(m,n)$ corresponds to a summation over $SL(2, \mathbb Z)$ images of worldsheet instantons, which arise from F-strings wrapping 2-cycles. This ensures inclusion of Euclidean D1-branes and bound states, $(p,q)$-strings, in the instanton corrections. Generically then, at large $|b_i|$ this correction is a function of all the $b_i$-axions present.

On general grounds then (see also the short discussion in~\cite{Grimm:2007xm} on p. 9) we expect the presence of a qualitatively analogous series of corrections which depend on the $c^i$-axions at large $|c^i|$ and which are exponentially suppressed in the volumes of the curves in $H_2^{+}$. As these terms will again contain a summation over all images under $SL(2, \mathbb Z)$, the individual terms will not be $\sim\cos(c^i)$ for a given $c^i$ but instead should have a dependence $\sim \cos(\alpha(n,m,\tau)_i c^i)$.
 
We may thus proceed by assuming the corrected K\"ahler potential to be of the form
\be\label{eq:instanton_corrections}
\mathcal{K}  =  - 2 \log \left\{{\mathcal V}_E  - e^{-\frac{2\pi v_+
}{\sqrt{g_s}}}\sum_{n,m} A_{n,m}(\tau,\bar\tau) \cos\left[\alpha(n,m,\tau,\bar\tau) c - \psi(m,n,\tau,\bar\tau,b)\right]\right\}\,,
\ee
for the case of one $c$-axion. The moduli potential often stabilizes $b$ due to its appearance in the K\"ahler potential already at the leading level~\cite{Grimm:2004uq,McAllister:2008hb}. Note, that this fixes $b$ not necessarily at zero which can give rise to a non-zero phase $\psi(m,n,\tau,\bar\tau,b)$.

This correction to the K\"ahler potential results in a proportional correction to the scalar potential of the theory, as well as corrections to the kinetic terms of  $\tau$ and $G^a$ which are exponentially suppressed in the curve volume $v_+$. Hence, normalization of the axion kinetic term, as e.g. in (5.19) of \cite{Flauger:2009ab}, will yield a correction to the potential of the form of eq.~\eqref{sumpot2}
\begin{equation}
 \delta V  \,\, \sim \, \, e^{-2\pi v_+/\sqrt{g_s}} \sum_i \tilde A_i \cos(\tilde\alpha_i \frac{\phi}{f} - \psi_i)\,,
\end{equation}
where we renamed the corresponding coefficients in eq.~\eqref{eq:instanton_corrections} as effective parameters $\tilde A_i, \tilde\alpha_i$.

For a very crude estimate of the number of terms one may expect to contribute to $\delta V$ we note that already values of $n,m$ smaller than 3 give ${\mathcal O}(10)$ possible terms in the $SL(2, \mathbb Z)$ sum, each of which can have contributions from the sum over rational curves in $H_2^{-}$. This could easily give the ${\cal O}(20)$ terms required in the sum of resonant bispectra to reproduce an equilateral-type non-Gaussian signal. 

Thus we have argued it to be plausible that a series of sinusoidal corrections with suitably varying frequencies may be achievable in connection with an axion monodromy generated potential, using the instanton corrections already considered at leading order in previous work. While the necessary and much deeper analysis of such constructions is left to the future, such a setup could form the starting point for a concrete model realizing the mechanism described in the preceding sections.

\section{Conclusions}
We have seen that a potential modulated by a sum of small oscillating terms 
will lead to a bispectrum given by a sum of resonant bispectrum terms which can sum to a bispectrum which closely approximates the equilateral bispectrum. This has potentially relevant implications for the aim of differentiating between inflationary models using data on non-Gaussianities.  This potential differentiating power rests on (a) non-Gaussianity being one of the few possible consequences of inflation that is not generic, (b) different classes of inflationary theories giving rise to different dominant bispectrum shapes, and (c) new data allowing us to set bounds on the amplitude of different shapes in the bispectrum being expected in the next few years \cite{Ade:2011ah}. 

In broad terms, single scalar field inflationary models with canonical kinetic terms have non-Gaussian signal of order the slow-roll parameters, so that they would be ruled out by any measurable detection of non-Gaussianity.  $P(X, \phi)$ theories (or more generally, Horndeski theories) with non-canonical kinetic terms of a single inflaton field give rise to predominantly equilateral-type non-Gaussianity \cite{Chen:2006nt}, while multi-field inflation gives rise to predominantly local-type non-Gaussianity and folded-type non-Gaussianity could indicate a non-Bunch Davies vacuum (For reviews see \cite{Chen:2010xka, Koyama:2010xj} and references therein). Orthogonal-type non-Gaussianity, found in \cite{Senatore:2009gt}, can be the dominant type in multi-field DBI galileon inflation with an induced gravity term \cite{RenauxPetel:2011dv, RenauxPetel:2011uk}. Different contributions to the non-Gaussianity can also appear in combination, as in \cite{Langlois:2008qf}. 

Here we have focussed on the possible degeneracy of equilateral-type non-Gaussianity, associated with single field theories with non-canonical kinetic terms, and summed resonant non-Gaussianity in a theory with canonical kinetic terms, arising from a potential modulated by a series of sinusoidal terms. It has been noted that within the class of non-canonical theories described by the Horndeski action, it may be difficult to differentiate between different models because some linear combination of the same shapes is always present \cite{Gao:2011qe, DeFelice:2011uc, RenauxPetel:2011sb}. The best way to differentiate between these models is therefore via conditions on the relative amplitudes of these terms which will differ for different models \cite{Ribeiro:2011ax}. Despite this, one might hope to be able to identify the presence of non-canonical kinetic terms, via the dependence of the equilateral signal on the reduced sound speed $f_{NL}^{eq} \sim \frac{1}{c_s^2}$.  

However, there is considerable degeneracy between observables even at this level.  A dominantly equilateral signal has been found in multi-field models such as trapped inflation \cite{Green:2009ds}, where there is no reduction in the speed of sound.  In addition, it was shown recently that $f_{NL}^{eq}$ may not depend explicitly on $c_s^2$ in a weakly coupled extension of the EFT of inflation with reduced sound speed, depending instead on the Hubble scale and the UV cutoff of the theory \cite{Gwyn:2012mw}. For axionic fields, a coupling $\phi F \tilde F$ to gauge fields is generically present, leading to the production of gauge quanta which can decay to give fluctuations in the inflaton density, giving rise to potentially large equilateral non-Gaussianity on top of the resonant-type signal \cite{Barnaby:2010vf, Barnaby:2011qe}. Our result adds to the existing degeneracies as we have shown that  even differentiating between canonical single field inflation models, with features in the potential and no 
coupling to gauge fields, and non-canonical single field models may be more difficult than initially anticipated.  

In the case where non-Gaussianity arises from simply broken shift symmetry of the action, the amount of resonantly generated equilateral-type non-Gaussianity allowed is subject to tight constraints which arise from the non-observation of oscillating contributions to the 2-point function, and the 
structure of the resonant $N$-point functions. These imply that detection of equilateral non-Gaussianity at a level greater than the PLANCK sensitivity of $f_{NL}\sim {\cal O}(5)$ will rule out a resonant origin~\cite{Behbahani:2011it}. In case a detection of non-Gaussianity eludes PLANCK, future 21cm observations may provide a detection capability of $f_{NL}\gtrsim {\cal O}(0.01)$ down to the slow-roll level itself due to the extremely large number of modes (${\cal O}(10^{16})$) available~\cite{Loeb:2003ya,Cooray:2006km}. For equilateral-type non-Gaussianities of this magnitude there would then be a degeneracy between a possible non-canonical and a resonant origin. However, if the shift symmetry is collectively broken, the magnitude of the non-Gaussian signal is no longer constrained by the bounds on the power spectrum, because breaking of scale invariance must involve a product of all the couplings necessary, so that $N$-point functions are no longer hierarchically suppressed with $N$~\cite{Behbahani:2012be}. In this case a large $f_{NL}$ with equilateral characteristics can be resonantly generated, as we have discussed. 

Let us note in closing that we have consciously kept most of discussion at the level of effective field theory. Beyond the very general considerations laid out in Section 4 any concrete analysis of embedding quasi-equilateral non-Gaussianity from multiple resonant oscillatory contributions into string theory e.g. along the lines of axion monodromy has to await much fuller treatment in the future.

\section*{Acknowledgments}
The authors are grateful to N.~Barnaby, D.~Green and E.~Pajer for several crucial and illuminating comments.
The work of A.W. was supported by the Impuls und Vernetzungsfond of the Helmholtz Association of German Research Centres under grant HZ-NG-603, and German Science Foundation (DFG) within the Collaborative Research Center 676 ``Particles, Strings and the Early Universe". R.G. is grateful for support by the European Research Council via the Starting Grant numbered 256994. R.G. was also supported during the initial stages of this work by an SFB fellowship within the Collaborative Research Center 676 ``Particles, Strings and the Early Universe'' and would like to thank the theory groups at DESY and the University of Hamburg for their hospitality at this time.

\appendix

\section{Computing the mode functions and bispectrum}
\label{computation}
\subsection{The mode functions}
\label{modefunctions}
We work in the comoving or unitary gauge where the inflationary perturbations are set to zero and spatial perturbations of the metric are written 
\be
\delta g_{ij}({\bf x}, t) = 2 a(t)^2 \mathcal R({\bf x}, t) \delta_{ij}.
\ee
The Fourier transform $ \mathcal R(k,t)$ can be written in terms of modes 
\be
\mathcal R({\bf k},t) = \mathcal R_k(t) a({\bf k}) + \mathcal R_k^\star (t) a^\dag (-{\bf k}),
\ee
for which the mode function $\mathcal R_k (t)$ satisfies the Mukhanov-Sasaki equation \cite{Mukhanov:1985rz, Sasaki:1986hm}, which for small $\epsilon$ is given by \cite{Weinberg:2008zzc}
\be
\frac{d^2 \mathcal R_k} {d x^2} - \frac{2(1 +2 \epsilon + \delta)}{x} \frac{d \mathcal R_k }{dx} + \mathcal R_k = 0,
\ee
where $ x = - k \tau$, for $ \tau$ the conformal time. 
For $x \ll 1$, $\mathcal R_k(t)$ approaches a constant value $\mathcal R_k^{(0)}$ which is related to the primordial scalar power spectrum. In the slow-roll approximation the Mukhanov-Sasaki equation is easily solved to give the answer in (2.16) of \cite{Flauger:2010ja}. But in our case, as in \cite{Flauger:2010ja}, $\dot \delta/H$ is large compared to $\epsilon$ and $\delta$ and this SR approximation is invalid. Neglecting $\epsilon_0$ and $\delta_0$ and to leading order in $\epsilon_\star$, the Mukhanov-Sasaki equation becomes
\be
\frac{d^2 \mathcal R_k} {d x^2} - \frac{2(1 +\delta_1(x))}{x} \frac{d \mathcal R_k }{dx} + \mathcal R_k = 0.
\ee
for 
\be
\delta_1 = - 3 \sum_i b_i^\star \sin \left (\frac{\phi_0 + c_i}{f_i} \right ).
\ee

 As in the single modulation case, the effect of $\delta_1$ is negligible at early times, for $ x\gg x_{{\rm res}, i} \, \, \forall i$, and 
\be
\mathcal R_k(x) = \mathcal R_{k,0}^{(o)} i \sqrt{\frac{\pi}{2}}x^{3/2} H_{3/2}^{(1)} (x),
\ee
where $\mathcal R_{k,0}^{(o)}$ is the value of $\mathcal R_k (x)$ outside the horizon in the absence of modulations and $H_{3/2}^{(1)} (x)$ is a Hankel function. Similarly for  $ x\ll x_{res, i}$, i.e. at sufficiently late times, the effect of $\delta_1$ is again negligible and the solution must be of the form
\be
\mathcal R_k (x) = \mathcal R_{k,0}^{(o)} \left [c_k^{(+)} i \sqrt{\frac{\pi}{2}} x^{3/2} H_{3/2}^{(1)} (x) - c_k^{(-)} i \sqrt{\frac{\pi}{2}} x^{3/2} H_{3/2}^{(2)} (x)  \right ] .
\ee
As in \cite{Flauger:2009ab}, $c_{k,i}^{(+)} = 1 + \mathcal O((b^i_\star)^2)$ at late times, where $c_k^{(+)} = \sum_i c_{k,i}^{(+)} = 1 + g(x)$ in the notation of \cite{Flauger:2009ab}. (One can see this by solving (3.31) of \cite{Flauger:2009ab} for each $i$).

Thus we look for a solution of the form 
\be
\mathcal R_k (x) = \mathcal R_{k,0}^{(o)} \left [i \sqrt{\frac{\pi}{2}} x^{3/2} H_{3/2}^{(1)} (x) - c_k^{(-)}(x) i \sqrt{\frac{\pi}{2}} x^{3/2} H_{3/2}^{(2)} (x)  \right ] ,
\ee
where $c_k^{(-)}(x)$ vanishes at early times. Then the Mukhanov-Sasaki equation gives an equation governing the time evolution of $c_k^{(-)}(x)$ which is given by (for $f_i \ll \sqrt{2 \epsilon_\star}$ for all $i$)
\be
\label{eqnforcminus}
\frac{d}{dx} \left [e^{-2 i x} \frac{d}{dx}  c_k^{(-)}(x) \right ]  =  - 2 i \frac{\delta_1(x)}{x}.
\ee

This can be written as
\bea
\frac{d}{dx} \left [e^{-2 i x} \frac{d}{dx}  c_k^{(-)}(x) \right ] & = & \frac{6 i}{x} \sum_i b_i^\star \sin (\frac{\phi_k}{f_i} + \frac{c_i}{f_i} + \frac{\sqrt{2 \epsilon_\star}}{f_i} \ln x)\,,\\
\frac{d}{dx} A(x) & = & \frac{6 i}{x} \sum_i b_i^\star \sin (\frac{\phi_k}{f_i} + \frac{c_i}{f_i} + \frac{\sqrt{2 \epsilon_\star}}{f_i} \ln x).
\eea

Let $A(x) = \sum_i A_i (x)$, so that 
\bea
\sum_i \frac{d}{dx} A_i(x) &=& \frac{6 i}{x} \sum_i b_i^\star \sin (\frac{\phi_k}{f_i} + \frac{c_i}{f_i} + \frac{\sqrt{2 \epsilon_\star}}{f_i} \ln x)\\
A_i(x) & = & - \frac{6 i b_i^\star f_i}{\sqrt{2 \epsilon_\star}} \cos (\frac{\phi_k}{f_i} + \frac{c_i}{f_i} + \frac{\sqrt{2 \epsilon_\star}}{f_i} \ln x ) 
\eea
is a solution. Then write
\be
A(x) = \sum_i A_i (x) = e^{-2 i x} \frac{d}{dx} c_k^{(-)} (x) = e^{-2 i x} \frac{d}{dx} \sum_i c_{k,i}^{(-)} (x)\,.
\ee
Then we find
\be
\label{ckminusderiv2}\frac{d}{dx} c_{k,i}^{(-)} (x) = - \frac{6 i b_i^\star f_i}{\sqrt{2 \epsilon_\star}} e^{2 i x} \cos  (\frac{\phi_k}{f_i} + \frac{c_i}{f_i} + \frac{\sqrt{2 \epsilon_\star}}{f_i} \ln x ). 
\ee

To find the mode functions outside the horizon, we use the expansions of the Hankel functions for small arguments:
\bea
H_{3/2}^{(1)} (x) &=& \frac{4}{3\sqrt{\pi}} \left ( \frac{x}{2}\right ) ^{3/2} - \frac{i }{2 \sqrt{\pi}} \left (\frac{2}{x} \right ) ^{3/2}\\
H_{3/2}^{(2)}(x) &= &\frac{4}{3 \sqrt{\pi}} \left ( \frac{x}{2}\right )^{3/2} + \frac{i}{2\sqrt{\pi}} \left (\frac{2}{x} \right ) ^{3/2}.
\eea
We find
\begin{align}
\begin{aligned}
\mathcal R_k (x) = &R_{k,0}^{(0)} \left [ 1 + \sum_i 3 b_i^\star \sqrt{\frac{f_i \pi}{2\sqrt{2 \epsilon_\star}}}  \cos \left (\frac{\phi_k + c_i}{f_i} \right ) \right.\\  &- \left. i \sum_i 3 b_i^\star \sqrt{\frac{f_i \pi}{2\sqrt{2 \epsilon_\star}}}  \sin \left (\frac{\phi_k + c_i}{f_i} \right )  + \mathcal O({x^3})\right ]\,,
\end{aligned}
\end{align}
and
\begin{equation}
 \label{mode_function_late2} |\mathcal R_k^{(0)} |^2 =  |\mathcal R_{k,0}^{(0)}|^2 \left [1 + \sum_i 3 b_i^\star \left (\frac{2 f_i \pi}{\sqrt{2 \epsilon_\star}} \right )^{1/2} \cos \left (\frac{\phi_k + c_i}{f_i}\right) \right ]. 
\end{equation}


\subsection{Mode functions in the bispectrum}
\label{unperturbed}
To see that we can use the unperturbed mode functions to calculate the bispectrum in Section~\ref{thebispectrum}, note first that at late times, for $x \ll1$, the mode function in eq.~\eqref{mode_function_late2} approaches a constant, given by $\mathcal R_{k,0}^{(o)} [1 + c_k^{(-)} ]$. The contribution at early times, $x \gg 1 $, within the horizon, is
\begin{align}
\begin{aligned}
\mathcal R_k (x) & =  \mathcal R_{k,0}^{(0)} \left [ - i x e^{i x} + i c_k^{(-)} (x) x e^{- i x}\right ]\\
& =  \mathcal R_{k,0} + \mathcal R_{k,1}, 
\end{aligned}
\end{align}
where we have used the $x \gg 1$ limit of the Hankel functions \cite{Flauger:2010ja}:
\begin{align}
\begin{aligned}
i \sqrt{\frac{\pi}{2}} x^{3/2} H_{3/2}^{(1)} (x) &\approx   - i x e^{i x}\\
i \sqrt{\frac{\pi}{2}} x^{3/2} H_{3/2}^{(2)} (x) &\approx - i x e^{-i x}
\end{aligned}
\end{align}
and $\mathcal R_{k,0}$ is the unperturbed part of the mode function. We find
\bea
\dot {\mathcal R_{k,0}} &=&  \mathcal R_{k,0}^{(0)}  x e^{i x}\\
\dot {\mathcal R_{k,1}} & = &  \mathcal R_{k,0}^{(0)} x e^{- i x}\left [ c_k^{(-)} (x) + i \frac{d}{dx}  c_k^{(-)} (x) \right ]\\
\Rightarrow \frac{|\dot {\mathcal R_{k,1}}|}{|\dot {\mathcal R_{k,0|}}|} & = & \left | c_k^{(-)} (x) + i \frac{d}{dx}  c_k^{(-)} (x)\right |.
\eea

 By eq.~\eqref{ckminus}) and eq.~\eqref{ckminusderiv2} we see that this ratio is small for $f_i \ll \sqrt{2 \epsilon_\star}$ (which is when the non-Gaussianity is appreciable). Thus the correction to the time derivative of the mode function is also small, and we can use the unperturbed mode functions and time derivatives in evaluating the three-point function.

\subsection{Squeezed limit consistency relation}
\label{consistency}
In~\cite{Flauger:2010ja} the squeezed limit consistency relation found in~\cite{Maldacena:2002vr,Creminelli:2004yq} was checked for resonant non-Gaussianity. It is straightforward to extend this to summed non-Gaussianities. The consistency condition can be phrased as~\cite{Flauger:2010ja}
\begin{equation}
 \lim_{k_3 \to 0} \langle \mathcal{R}(\boldsymbol{k_1},t) \mathcal{R}(\boldsymbol{k_2},t) \mathcal{R}(\boldsymbol{k_3},t) \rangle \simeq - (2 \pi)^3 \delta^3(\boldsymbol{k_1} + \boldsymbol{k_2} + \boldsymbol{k_3}) |\mathcal{R}^{(0)}_{k_3}|^2 |\mathcal{R}^{(0)}_{k}|^2 \frac{d\, \ln \Delta^2_{\mathcal{R}}(k)}{d\, \ln k}\,, \label{squeezedcons}
\end{equation}
where in the squeezed limit $k_1 \simeq k_2 = k \gg k_3$. Furthermore, $|\mathcal{R}^{(0)}_{k}|^2 = 2 \pi^2 \Delta_{\mathcal{R}}^2 / k^3$, with\footnote{For this discussion we set the phases $c_i = 0$.}
\begin{equation}
 \Delta_{\mathcal{R}}^2(k) = \Delta_{\mathcal{R}}^2(k_\star) \left(\frac{k}{k_\star} \right)^{n_s-1} \left[ 1 + \sum_i 3 b_i^\star \sqrt{2\pi} \left(\frac{f_i}{\sqrt{2\epsilon_\star}} \right)^{1/2} \cos \left( \frac{\sqrt{2\epsilon_\star}}{f_i} \ln \frac{k}{k_\star} \right)\right]\,,
\end{equation}
such that
\begin{equation}
 \frac{d\, \ln \Delta^2_{\mathcal{R}}(k)}{d\, \ln k} \simeq - \sum_i 3 b_i^\star \sqrt{2\pi} \left(\frac{\sqrt{2\epsilon_\star}}{f_i} \right)^{1/2} \sin \left( \frac{\sqrt{2\epsilon_\star}}{f_i} \ln \frac{k}{k_\star} \right)\,,
\end{equation}
neglecting slow roll corrections to $n_s = 1$. Now we can write
\begin{align}
 \begin{aligned}
  \lim_{k_3 \to 0} \langle \mathcal{R}(\boldsymbol{k_1},t) \mathcal{R}(\boldsymbol{k_2},t) \mathcal{R}(\boldsymbol{k_3},t) \rangle \simeq \,\,& (2 \pi)^7 \Delta^4_{\mathcal{R}} \frac{1}{k_1 k_2 k_3} \delta^3(\boldsymbol{k_1} + \boldsymbol{k_2} + \boldsymbol{k_3})\\ &\times \frac{2k}{k_3 } \sum_i \frac{3 b_i^\star \sqrt{2\pi}}{8} \left(\frac{\sqrt{2\epsilon_\star}}{f_i} \right)^{1/2} \sin \left( \frac{\sqrt{2\epsilon_\star}}{f_i} \ln \frac{k}{k_\star} \right)\,.
 \end{aligned}
\end{align}
Comparing with eq.~\eqref{threepointGgen}, to fulfill the consistency condition eq.~\eqref{squeezedcons}, $\mathcal G_{res}$ has to be given as
\begin{equation}
 \frac{\mathcal G_{res}(k,k,k_3)}{k^2 k_3} \simeq \frac{2k}{k_3 } \sum_i \frac{3 b_i^\star \sqrt{2\pi}}{8} \left(\frac{\sqrt{2\epsilon_\star}}{f_i} \right)^{1/2} \sin \left( \frac{\sqrt{2\epsilon_\star}}{f_i} \ln \frac{k}{k_\star} \right)\,,
\end{equation}
in the squeezed limit. This is indeed the case up to a phase, as can be seen by taking the limit $k_3 \to 0$ of eq.~\eqref{sumbispectrum}.

\bibliographystyle{JHEP.bst}
\bibliography{sumresonant}
\end{document}